\documentclass[conference]{IEEEtran}
\IEEEoverridecommandlockouts
\usepackage{cite}
\usepackage{amsmath,amssymb,amsfonts}
\usepackage{algorithmic}
\usepackage{graphicx}
\usepackage{textcomp}
\usepackage{xcolor}
\usepackage{verbatim}
\usepackage{subfigure}
\def\BibTeX{{\rm B\kern-.05em{\sc i\kern-.025em b}\kern-.08em
    T\kern-.1667em\lower.7ex\hbox{E}\kern-.125emX}}

\begin{document}

\title{Modulation and Classification of Mixed Signals Based on Deep Learning\\
}


\author{Jiahao Xu, Zihuai Lin\\
	School of Electrical and Information Engineering, The University of 
	Sydney, Australia\\
	Emails:	zihuai.lin@sydney.edu.au. 
}

\maketitle

\begin{abstract}
\noindent Modulation classification technology has important application value in both commercial and military fields. Modulation classification techniques can be divided into two types, one is based on likelihood ratio and another is based on signal characteristics. Because the modulation classification technology based on the likelihood ratio is too complex to calculate, the modulation classification technology based on signal characteristics is usually more widely used. The traditional modulation classification technology based on signal characteristics first designs and extracts signal characteristics, and then designs classification rules to classify signal characteristics by modulation format. Manually designed signal characteristics and classification rules usually have limited recognition accuracy in complex channel environments. Therefore, modulation classification techniques based on deep learning are proposed to automatically learn classification rules from feature data to improve classification accuracy. With the rapid development of information nowadays, spectrum resources are becoming more and more scarce, leading to a shift in the research direction from the modulation classification of a single signal to the modulation classification of multiple signals on the same channel. Therefore, the emergence of an effective mixed signals automatic modulation classification technology have important significance.
Considering that NOMA technology has deeper requirements for the modulation classification of mixed signals under different power, this paper mainly introduces and uses a variety of deep learning networks to classify such mixed signals. First, the modulation classification of a single signal based on the existing CNN model is reproduced. We then develop new methods to improve the basic CNN structure and apply it to the modulation classification of mixed signals. Meanwhile, the effects of the number of training sets, the type of training sets and the training methods on the recognition accuracy of mixed signals are studied. 
Second, we investigate some deep learning models based on CNN (ResNet34, hierarchical structure) and other deep learning models (LSTM, CLDNN). It can be seen although the time and space complexity of these algorithms have increased, different deep learning models have different effects on the modulation classification problem of mixed signals at different power. Generally speaking, higher accuracy gains can be achieved.
\end{abstract}


\section{Introduction}

Modulation classification technology is one of the technologies of cognitive radio in the communication field. Modulation identification is used to determine the modulation format used by the signal based on the signal received by the receiver. Modulation classification technology has both practical value and theoretical significance.

With the rapid development of information nowadays, spectrum resources are becoming more and more scarce, leading to a shift in the research direction from the modulation classification of a single signal to the modulation classification of multiple signals on the same channel. Take non-orthogonal multiple access technology (NOMA) as an example. With the rapid development of 5G, this technology has been applied to actual businesses. One of the cores of this technology is the multiplexing of power. The maximum performance gain of the system is achieved by allocating signals of different energy to different users. But this brings great challenges to the signal recognition at the receiving end. Traditional signal processing methods cannot filter and separate signals in the time or frequency domain. Therefore, the emergence of an effective mixed signals automatic modulation classification technology will have important significance.

Recently, deep learning has demonstrated strong performance in image recognition, speech classification, and many other areas. Its ability to extract local features based on neural network modeling has also been applied to the processing of communication signals. However, deep learning is now being used by the majority of individuals to achieve automatic modulation classification of a single signal. Theoretical findings of automatic modulation classification are scarce in the situation of a single channel and multiple signals. As a result, investigating the classification and identification of modulated signals with different power under the same channel is of great significance.

This paper is mainly based on the modulation recognition technology of neural networks to realize automatic recognition and classification of mixed signals. 
First, the traditional CNN is used as a feature extractor and a classifier simultaneously to identify and classify the features of a single signal. And further finding out the influence of the diversity of the training set on the performance of CNN modulation recognition. Comprehensive realization of the recurrence of known results.
Then, CNN architecture is improved. Based on the requirements of NOMA technology, the mixed signals with different power under the same channel are identified and classified. The modulation recognition accuracy under different SNRs is improved by changing the number of training sets and the training mode with the goal of achieving the upper limit computed by the Average Likelihood Ratio Test (ALRT) algorithm. 
Finally, considering that traditional CNN's recognition accuracy for mixed signals is not very ideal. Use new CNN networks (ResNet, Hierarchical Structure), RNN networks, and other deep learning models to try to raise the accuracy of mixed signals recognition.

The main contributions of this paper are as follows: (1) the traditional CNN-based single-signal modulation classification model is analyzed, simulated, and reproduced; (2) propose and develop deep learning-based classification and recognition of mixed signals of varying power within the same channel; (3) explore the differences in classification accuracy in different scenarios by changing the number of training samples, training modes, and signal power; (4) employ a variety of deep learning models (CNN, ResNet, LSTM, CLDNN) to achieve the modulation classification of mixed signals, and identify the deep learning model with the highest recognition rate as much as possible.


The outline of this paper is given below. 
Section I mainly introduces the research significance of the problem researched in this paper, the current research status of the problem, and the research content of this paper.
Section II first introduces the NOMA technology, which leads to the research of the classification of mixed signals under different power. Based on this problem, a variety of deep learning models including CNN, ResNet, CLDNN, LSTM, and hierarchical structures are introduced. Then the maximum likelihood estimation algorithm, i.e., the ALRT algorithm is introduced as the upper limit of the accuracy of the signal classification problem.
Section III first gives a particular description of the single signal model and mixed signals models used in the thesis, highlighting the advantages of symbol sequences in signal recognition problems. At the same time, it specifically introduces the design details of several deep learning models mentioned in section II and related design methods mainly used in the paper.
Section IV presents the simulation results. Section V summarise the work.


\section{Background}

In this section, we will review three main aspects of the technique: (1) the development of automatic modulation classification technology; (2) the research status of deep learning for single signal modulation classification; (3) some theories of deep learning on mixed signals automatic modulation recognition.

\subsection{The Development of Automatic Modulation Classification Technology}
\begin{figure}[htbp]   
    \centering  
    \includegraphics[width=8.5cm]{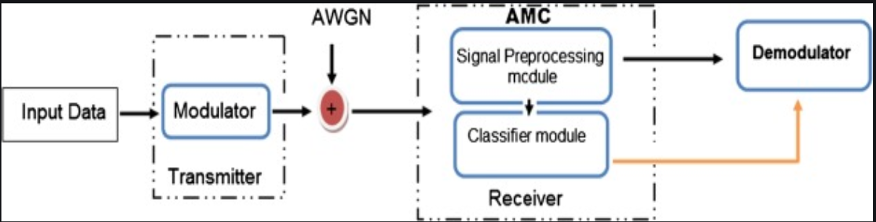}
    \caption{AMC framework}    
    \label{AMC framework}  
\end{figure}

Automatic modulation classification is a very important link between the channel and the demodulation system. If the receiver does not understand the signal or knows less, this will be an essential identification process. The automatic modulation classifier generally includes two design processes: preprocessing the input signal and selecting the classification method as shown in Fig. \ref{AMC framework}. The current classification method includes two categories: based on likelihood (LB) and based on feature (FB)\cite{dobre2007survey}.

\textbf{LB method}: The authors of \cite{wei2000maximum} first utilized the maximum likelihood method for the classification of digital signal orthogonal modulation. They proved that the IQ domain data of the signal can be applied to the modulation classification of the signal. And when the amount of data is infinite, the classification effect is close to 0 error. In \cite{chavali2011maximum}, the researchers proposed an algorithm for classifying digital amplitude and phase-modulated signals. This algorithm is suitable for fading channels with mixed Gaussian noise. Mixed Gaussian noise appears in most practical channels. This algorithm uses the method of expectation maximization to estimate the parameters of these noises and classify the overall signal by a mixed likelihood ratio.

\begin{figure}[htbp]   
    \centering  
    \includegraphics[width=6cm]{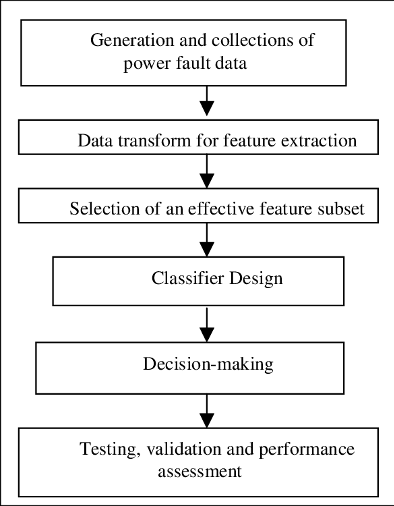}
    \caption{Feature-based method}     
    \label{Feature-based method}  
\end{figure}

\textbf{FB method}: Since the actual application cannot obtain a large amount of required information through the LB method, the FB method is the focus of the researchers in recent years. The core components of FB are feature extraction and classifier. The specific steps were shown in Fig. \ref{Feature-based method} . Various types of feature values have been extracted and applied to the AMC algorithm. For example, the authors of \cite{azzouz1995automatic} extracted instantaneous features from the instantaneous amplitude and phase expressions of digital modulation signals. The higher-order statistical moments (transformation-based features) in \cite{liu2012novel} were extracted from Fourier transform and wavelet transform. In \cite{das2016cumulant}, the characteristic value of the high-order cumulative quantity (HOC) was obtained from the difference of the cumulative quantity of received signals. The high-order cumulant feature can also solve the additional interference caused by Gaussian white noise, and is suitable for modulation classification under low to medium signal-to-noise ratio.

There are two main categories in the realization of classifiers, linear classifiers and nonlinear classifiers. The decision tree in the linear classifier is one of the most widely used linear classifiers \cite{azzouz1995automatic}. The decision tree can handle large-scale data very well, but there is no good way to deal with linear indivisible problems. For non-linear classifiers, there are mainly neural networks and SVM. The neural network processing method was proposed in 2009 \cite{hassan2009automatic} and was investigated in the subsequent literature. The SVM algorithm implicitly mapped its input into a high-dimensional feature space through kernel techniques to achieve classification \cite{orlic2010multipath}. The disadvantage of the FB method is that the parameters must be manually adjusted to adapt to the channel and modulation environment. This causes a lot of extra resources to be wasted, so deep learning can be used to solve this problem.

\subsection{Deep Learning Method for Single Signal Modulation Classification} 
\begin{figure}[htbp]   
    \centering  
    \includegraphics[width=9cm]{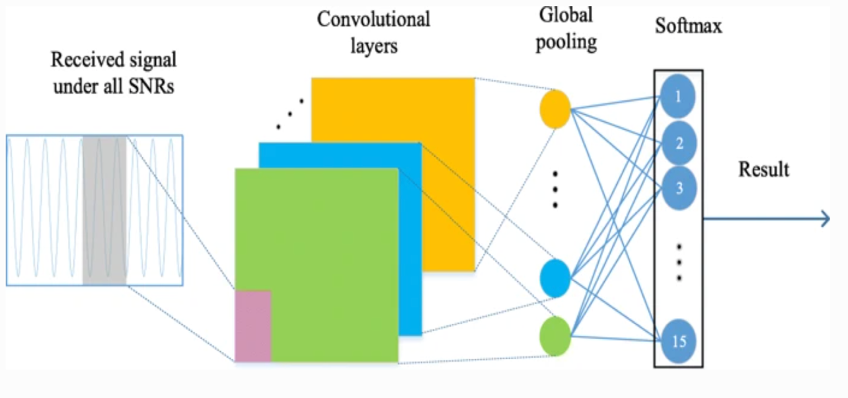}
    \caption{Signal modulation classification using CNN}    %
    \label{Signal modulation classification using CNN}  %
\end{figure}

So as to improve the accuracy and efficiency of single signal recognition, the way of deep learning has been used for the modulation classification of the single signal. The most common one is the network frame using CNN shown in Fig. \ref{Signal modulation classification using CNN} to modulate and recognize a single signal.

This CNN-based single-signal recognition method was proposed in \cite{o2016convolutional} in 2016. It uses the time-domain characteristics of complex signals for feature extraction and compares it with expert feature-based methods. It is found that this method not only improves the performance but also proves that this blind time learning method can be applied to densely coded time series. In addition to the deep learning model of CNN, others also use various methods to realize the automatic recognition and modulation of single signals. A heterogeneous depth model fusion method to solve such problems was proposed in \cite{zhang2018automatic1}. This method obtains a high-performance filter by fusing CNN and LSTM models and improves the classification accuracy. The characteristics of AMC indicate that the ability has been greatly improved. In \cite{zhang2018automatic2}, the data was first preprocessed, the original IQ data was then combined with the fourth-order cumulant (FOC) of the signal, and the IQ-FOC data representation form was proposed. It is more robust to noise and performs better than CNN network under LSTM network. Ref. \cite{mendis2016deep} adopted a deep belief network (DBN) to improve the accuracy of signal detection and classification in a high-noise environment through the use of spectral correlation functions (SCF). Ref. \cite{dai2016automatic} tried to use sparse autoencoders to extract features from the fuzzy function image of the signal. The feature values obtained in this way will be processed by the relevant regression classifier. The advantage of this processing method lies in the reliability of the signal classification and the classification accuracy in the case of the low SNR.

In short, the AMC method based on the single signal mainly concentrates on the preprocessing of the original signal data or the related improvement of the deep learning network. These methods have never been used in the AMC of mixed signals. Considering the complexity of the mixed signals, there needs to be an effective method to solve this mixed signals AMC problem.

\subsection{Deep Learning Method for Mixed Signals Modulation } 
\begin{figure}[htbp]   %
    \centering  %
    \includegraphics[width=9cm]{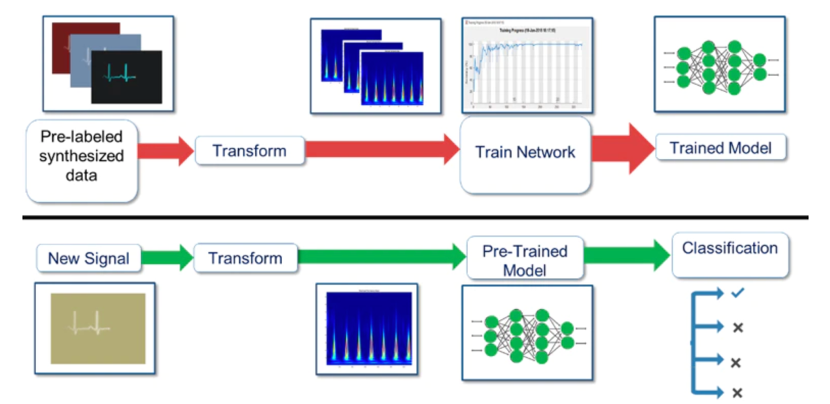}
    \caption{Deep learning structure for mixed signals AMC}    %
    \label{Deep learning structure for mixed signals AMC}  %
\end{figure}

The deep learning network is still in its infancy for automatic modulation recognition of mixed signals. There are not many documents describing the solution to this problem. Fig. \ref{Deep learning structure for mixed signals AMC} can be regarded as a deep learning network structure to solve this problem.
Judging from the few pieces of literatures on the classification of mixed signals automatic modulation, the main methods used are:

1. The CNN network is used to identify the IQ characteristics of the mixed signals. This method takes advantage of the independence of one-dimensional signal data and can automatically extract the characteristics of features through CNN networks, reducing the number of parameters, and improving the recognition efficiency \cite{yin2019co}.

2.The authors of \cite{sun2018automatic} remodeled the signal into a new image signal through oversampling and performed automatic modulation recognition through the CNN network. The recognition efficiency of this method will change with the setting of the oversampling rate value, but the recognition work can be completed well when the oversampling rate is large.

3. Modulation and classification of mixed signals are done by capsule network \cite{zhou2019blind}. Since the weight and output of the capsule network are vectors, it is more efficient for feature recognition. It can identify the characteristics of individual signals in the mixed signals and be directly processed by the network without additional work. The difficulty is that the complexity of the network is too high.

4. A new deep convolutional network demodulator (DCND) to demodulate mixed signals is investigated in \cite{lin2017deep}. In this way, the original data is coherently demodulated by combining two CNN networks. Effectively reduce the bit error rate. However, its demodulation performance on BASK, BFSK, and BPSK is not perfect. 

\section{Selection of Signal Models and Design of Deep Learning Models\label{cha:methods}}

The content of this section can be divided into two parts: The first part is the construction of the signal model. For the use of deep learning methods to solve signal classification problems, two signal types are usually used: the oversampling sequence of the baseband signal\cite{o2016radio,o2017semi,west2017deep} and the baseband symbol sequence \cite{hu2018robust,meng2018automatic}. Taking into account the need to mix signals under different power, the simplified symbol sequence can remove additional information while maintaining the original characteristics of the signal and be accepted by the deep learning model. Therefore, the signal type used in this paper is a symbol sequence. For the construction of a single signal, there exist many methods, such as, \cite{zhu2015automatic,swami2000hierarchical,peng2017modulation}. For mixed signals, the mixed signals model of the baseband symbol sequence can be obtained according to the model of the mixed baseband signals given in \cite{o2016convolutional}. It is worth noting that the difference in power will be emphasized in the mixed signals model.

The second part is the design of multiple deep learning models. This part mainly designs five models to classify mixed signals under different power. Among them, CNN is the main research focus, and different CNN models are used to classify and recognize single signals and mixed signals. In the design process, the specific structure of CNN and the setting of the data set for training are mainly considered. For ResNet, hierarchical structure, LSTM, and CLDNN models, the main design idea is to modify the existing models. And these models will be mainly used in the recognition and classification of mixed signals, so the setting of the training data set is consistent with CNN, and will not be repeated in the paper. All simulations and result analysis of these models will be discussed in Section \ref{cha:results}.

\subsection{Signal Model}
\subsubsection{Single Signal Model}
According to \cite{dobre2007survey}, the complex envelope of the baseband signal is:
\begin{equation}
r(t)=s\left(t ; \boldsymbol{u}_{k}\right)+g(t)
\end{equation}

Where $g(t)$ is the complex Gaussian white noise with a mean value of zero and a double-side band power spectral density of $N_0/ 2$, $s\left(t ; \boldsymbol{u}_{k}\right)$ is a comprehensive representation of the modulated digital signal without noise interference, which can be represented by:
\begin{equation}{\label{eq1}}
s\left(t ; u_{k}\right)=a e^{j\left(2 \pi \Delta f t+\theta_{c}\right)} \sum_{n=0}^{L-1} s_{n}^{k, i} e^{j \theta_{n}} v(t-n T-\epsilon T)
\end{equation}

In eq. \ref{eq1}, $a$ is the amplitude of the signal, $\theta_c$ represents the fixed phase offset caused by the initial phase of the carrier and the propagation delay, $N$ is the number of symbols in the observation interval, $s_n^{k,i}$ represents the $i_{th}$ constellation point under the $k_{th}$ modulation system, $\Delta f\ $ is the frequency offset caused by the down-conversion process, $\theta_n\ $ represents phase jitter, $T$ is the symbol period, $\epsilon$ represents symbol timing offset, $v (t)$ represents the combined effect of the channel influence $h (t)$ and the pulse shaping function $p (t)$.

Performing discrete modulation recognition processing on the time domain continuous signal $r(t)$ obtained above, the discrete point signal output after the matched filter can be obtained:
\begin{equation}
r(n)=e^{j\left(2 \pi f_{0} T_{n}+\theta_{n}\right)} \sum_{\ell=0}^{L-1} s(\ell) h(n T-\ell T-\epsilon T)+g(n)
\end{equation}
\\where $s\left(\ell\right)\ $ is the sending symbol sequence, $h(\bullet)$ is the channel response function, $T$ is the symbol interval, $\epsilon$ is synchronization error, $f_0$ represents the frequency offset, $\theta_n$ is the phase jitter, $g(n)$ is the noise.

The model is further simplified, assuming that the symbol interval is known, the signal is synchronized, and the phase jitter has been eliminated. A symbol sequence model with only frequency offset, channel response and noise interference can be obtained.
\begin{equation}
r(n)=e^{j 2 \pi f_{0} T_{n}} s(n) h(n)+g(n)
\end{equation}

It is worth noting that when deep learning models such as CNN are used as the complex symbol sequence of the input baseband signal for modulation recognition, the symbol sequence needs to be preprocessed. The reasons are: first, CNN's backpropagation-based training algorithm cannot handle complex input; Second, when using CNN to process pictures, the input picture usually has three dimensions: length, width and channel. This section uses the three-dimensional convolutional layer used in image processing, so it is necessary to expand the dimensions of the signal to fit the convolutional layer. The preprocessing process of the symbol sequence is as follows:
\begin{equation}
[r]^{1 * L}=\left[\begin{array}{l}
\Re(r) \\
\beth(r)
\end{array}\right]^{2 * L * 1}
\end{equation}

There are several advantages of using a simplified signal model based on symbol sequences: (1) Symbol sequences can be represented by constellation diagrams, which is convenient for the subsequent use of clustering-based modulation recognition algorithms; (2) There are many communication signal interference factors, and simplified signals are used. The model can theoretically analyze the feasibility of the algorithm in combination with the method of controlling variables, and better evaluate the performance of the algorithm.

The signal model used in this part mainly refers to the signal model in the literature \cite{zhu2015automatic,swami2000hierarchical,peng2017modulation}, and further research is carried out on this basis.

\subsubsection{Mixed Signals Model}

As described earlier, the symbolic signal model of a single signal can be written as

\begin{equation}
r_{i}(n)=\sqrt{E_{i}} e^{j\left(2 \pi f_{i} T_{n}+\theta_{n}\right)} \sum_{\ell=0}^{L-1} s(\ell) h(n T-\ell T-\epsilon T), 
\end{equation}
where $i\in\{1,2, \ldots, N\}$, $\sqrt{E_i}$ is the power of the signal,$s\left(\ell\right)$ is the sending symbol sequence, $h(\bullet)$ is the channel response function, $T$ is the symbol interval, $\epsilon$ is synchronization error, $f_i$ represents the frequency offset, $\theta_n$ is the phase jitter. 

\begin{figure}[htbp]   %
    \centering  %
    \includegraphics[width=9cm]{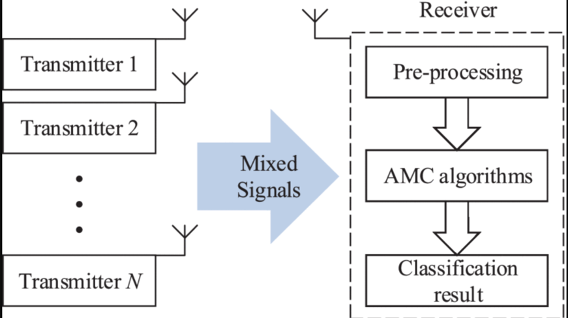}
    \caption{Multi-signal transmission structure}    %
    \label{Multi-signal transmission structure}  %
\end{figure}

In this part, the AMC task is conducted on two co-channel signals received by a single receiver. The process is depicted in Fig. \ref{Multi-signal transmission structure}.

Under the condition of two co-channel signals, the received signal $r(n)$ can be represented as:

\begin{equation}
r(n)=\sum_{i=1}^{2} r_{i}(n)+g(n)
\end{equation}
where $g(n)$ is a complex Gaussian baseband noise.
It should be noticed that in the signal model of this part, the mixing type of the signal and the difference of signal power will be considered.

\subsection{Design of Convolutional Neural Network}

\subsubsection{Design of CNN for Single Signal AMC}

The design of CNN mainly includes the design of network structure parameters and the design of network training parameters. The network structure parameters refer to the parameters related to the structure of the network, including the amount of convolutional layers, the size of the convolution kernel, the convolution step length, whether to use the bias term, the type of activation function, the initialization method, and so on. Network training parameters refer to parameters related to the training process of the network, including training data type, training data quantity, loss function, optimization method, learning rate, fuzzy factor, stopping conditions, etc. Some of the parameters have a significant impact on the performance of the trained CNN. For the modulation recognition scheme of a single signal, the 4-layer CNN network structure mentioned in \cite{o2016convolutional} will be adopted in this section. The specific design scheme is shown in Table \ref{table:CNN structure for single signal}.

\begin{table}[htbp]
\renewcommand\arraystretch{2}
\caption{CNN structure for single signal}
\label{table:CNN structure for single signal}
\setlength{\tabcolsep}{1mm}{
\begin{tabular}{|c|c|c|}
\hline
\textbf{Layer}        & \textbf{Parameter ($n * a * b$)} & \textbf{Activation Function} \\ \hline
Input layer           & /                  & /                            \\ \hline
Convolutional layer   & 64*2*4             & ReLU                         \\ \hline
Convolutional layer   & 16*1*4             & ReLU                         \\ \hline
Flattening layer      & /                  & /                            \\ \hline
Fully connected layer & 64                 & ReLU                         \\ \hline
Fully connected layer & 16                 & ReLU                         \\ \hline
Output layer          & 4                  & Softmax                      \\ \hline
\end{tabular}}
\end{table}

In Table \ref{table:CNN structure for single signal}, $n$ in the parameter $n * a * b $ represents the number of convolution kernels, and $a * b$ represents the size of the convolution kernel. ReLU is used as the activation function for the convolutional layer and the first two fully connected layers, and Softmax is used as the activation function for the output layer.

The classification signal set used is $\Omega_4=\{BPSK,4QAM,8PSK,16QAM \}$, signal sequence length $L = 100$, the $i_{th}$ received signal sequence is $r_i(n)=s_i(n)+g_i(n)$, where $n\in\{1,\cdots,L\}, g_{i}(n) \sim CN(0,1)$. The average SNR of each symbol sequence obeys a uniform distribution between $[-10, 20]$ dB, the power of the symbols within each symbol sequence is the same, and the noise is different. The training data adopts a Monte Carlo method to generate $N_{sample}$ symbol sequences for each modulation category, and a total of $4{\ast N}_{sample}$ signal sequences are generated. Among them, $0.1*N_{sample}$ data is used as the cross-validation set, and the rest is used as the training set. The test data uses the same method. The SNR test range is $[-10, 20]$ dB, and the test interval is 2dB. The symbol sequence of each modulation category corresponds to a specific SNR to generate 1000 test sets. Finally, the test results are averaged to obtain the modulation recognition accuracy rate.

It is worth noting that when using CNN to perform modulation recognition on the complex symbol sequence of the input baseband signal, the symbol sequence needs to be preprocessed. The reason is that: First, the training algorithm based on backpropagation that is limited by CNN cannot handle complex input; Second, when using CNN to process pictures, the input picture usually has three dimensions, namely length, width, and channel. This work uses the three-dimensional convolutional layer used in image processing, so the dimensions of the signal need to be expanded to fit the convolutional layer.

\subsubsection{Design of CNN for Mixed Signals AMC}

The first issue to be determined in this part is the choice of the data type. Unlike the signal set composition method of a single signal, the composition method of mixed signals is a pairwise combination of different modulation types, and the power difference of different signals must be considered. In this section, in order to better determine the overall CNN structure, the classification signal set is 
$\Omega_4=\{BPSK,4QAM,8PSK,16QAM\}$
so the total combination set is 
$\Omega_6=\{BPSK+4QAM,BPSK+8PSK,BPSK+16QAM,4QAM+8PSK,4QAM+16QAM, 8PSK+16QAM\}$. 

Power difference will not be considered in this part which means the power presented by different SNR will be divided equally to each modulation scheme. Other parameters will still be the same, signal sequence length $L = 100$, the $i_{th}$ received signal sequence satisfies $r_i(n)=s_i(n)+g_i(n)$, where $n=1,\cdots,L, g_{i}(n) \sim CN(0,1)$. The average SNR of each symbol sequence obeys a uniform distribution between $[-10, 20]$ dB and the noise is different. The training data adopts a Monte Carlo method to generate $N_{sample}$ symbol sequences for each modulation category, and a total of $6{\ast N}_{sample}$ signal sequences are generated. Among them, $0.1*N_{sample}$ data is utilized as the cross-validation set, and the rest is utilized as the training set. The test data uses the same method. The SNR test range is $[-10, 20]$ dB, and the test interval is 2dB. The symbol sequence of each modulation category corresponds to a specific SNR to generate 1000 test sets. Finally, the test results are averaged to obtain the modulation recognition accuracy rate.

The CNN model used in the single-signal modulation classification method shown in Table \ref{table:CNN structure for single signal}  will be first used in the mixed signals to test the accuracy of its classification in order to facilitate subsequent improvements.

\begin{figure}[htbp]   %
    \centering  %
    \includegraphics[width=9cm]{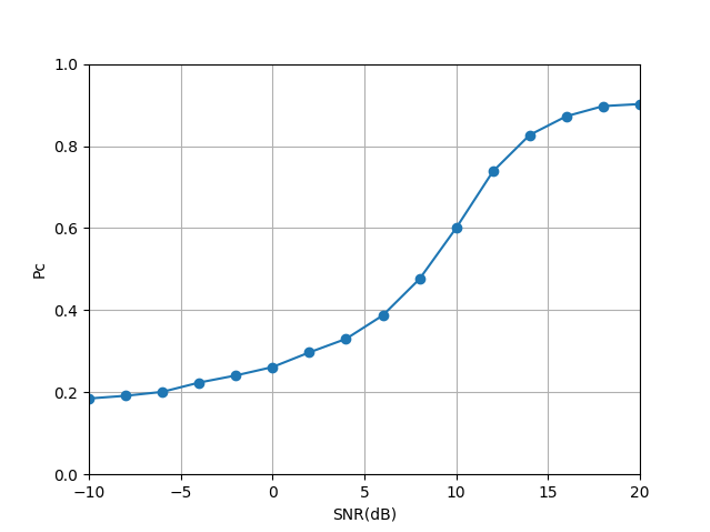}
    \caption{Mixed signals classification accuracy using CNN model }    %
    \label{Mixed signals classification accuracy using CNN model }  %
\end{figure}

Fig. \ref{Mixed signals classification accuracy using CNN model } shows that although the CNN network used in \cite{o2016convolutional} can classify mixed signals effectively, the comprehensive recognition accuracy under various SNRs is low. To better improve the classification performance of the CNN network, more convolutional layers, number of convolutional kernels, and fully connected layers will be used. At the same time, maximum pooling layers will also be added to reduce overfitting. The new CNN network suitable for mixed-signal modulation classification will be shown in Table \ref{table:CNN structure for mixed signals}.

\begin{table}[htbp]
\renewcommand\arraystretch{2}
\caption{CNN structure for mixed signals}
\label{table:CNN structure for mixed signals}
\setlength{\tabcolsep}{1mm}{
\begin{tabular}{|c|c|c|}
\hline
\textbf{Layer}        & \textbf{Parameter $(n * a * b)$} & \textbf{Activation Function} \\ \hline
Input layer           & /                  & /                            \\ \hline
Convolutional layer   & 64*2*2             & ReLU                         \\ \hline
Maximum pooling layer & 1*2                & /                            \\ \hline
Convolutional layer   & 32*1*4             & ReLU                         \\ \hline
Maximum pooling layer & 1*2                & /                            \\ \hline
Convolutional layer   & 20*1*4             & ReLU                         \\ \hline
Maximum pooling layer & 1*2                & /                            \\ \hline
Convolutional layer   & 16*1*4             & ReLU                         \\ \hline
Maximum pooling layer & 1*2                & /                            \\ \hline
Flattening layer      & /                  & /                            \\ \hline
Fully connected layer & 256                & ReLU                         \\ \hline
Fully connected layer & 64                 & ReLU                         \\ \hline
Fully connected layer & 16                 & ReLU                         \\ \hline
Output layer          & 6                  & Softmax                      \\ \hline
\end{tabular}}
\end{table}

In Table \ref{table:CNN structure for mixed signals},  $n$ in the parameter $n * a * b$ represents the number of convolution kernels, and $a * b$ represents the size of the convolution kernel. The parameter $c*d$ for the maximum pooling layer represents the pooling size of the pooling layer. ReLU is used as the activation function for the convolutional layer and the first three fully connected layers, and Softmax is used as the activation function for the output layer.

\subsection{Design of ResNet}

The design of the Resnet network structure follows two design rules: 1) The layers have the similar number of filters under the condition of having the same output feature map size; 2) To ensure the time complexity of each layer, the number of filters should be changed along with the feature map size.

\begin{figure}[htbp]   %
    \centering  %
    \includegraphics[width=9cm]{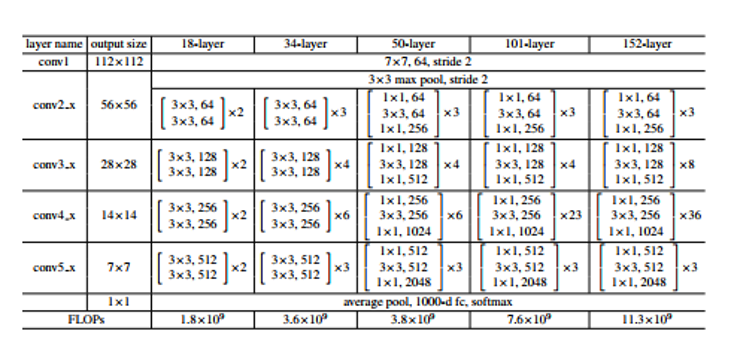}
    \caption{Common ResNet structure}    %
    \label{Common ResNet structure }  %
\end{figure}

Fig. \ref{Common ResNet structure } shows the five commonly used ResNet network structures. This type of deep learning method is mainly applied to the classification of mixed signals as an improvement of the CNN method. Taking into account the characteristics of the mixed signals model, the complexity of the operation and many other factors, here will not make too many changes to the ResNet model. Simultaneously, this part will use ResNet34 as the research object.

\subsection{Design of Hierarchical Structure}

For the hierarchical structure,  what deep learning solves is actually a multi-classification problem. Like many deep learning methods, CNN is a common method for mixed -signal classification, but this method usually requires a large-scale signal set. The problem faced by multi-label classification tasks is mainly the problem of data sparseness caused by multi-label. In multi-label classification, it is often encountered that the labels are parallel and have a hierarchical structure. Take the mixed-signal classification when the signal power ratio is 2:1 as an example. The entire signal category can be classified at the first level based on the stronger modulated signal, and then the weaker signal can be classified on the basis of the first level classification. 

\begin{figure}[htbp]  
    \centering  
    \includegraphics[width=9cm]{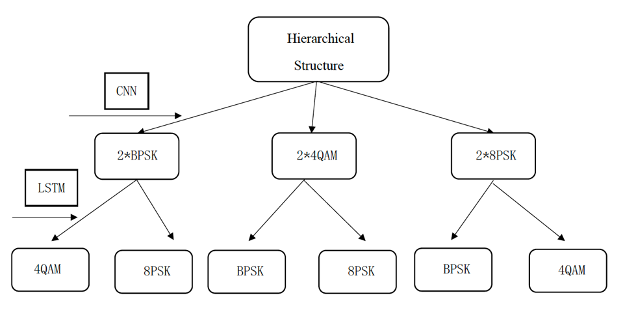}
    \caption{Hierarchical classification for mixed signals with different power}    %
    \label{Hierarchical classification for mixed signals with different power}  
\end{figure}

In this part, the basic CNN structure shown in Table \ref{table:CNN structure for mixed signals} will be used to first separate the stronger signal in the mixed signals, and then the LSTM method which will be described in the next section will classify and identify the weaker signals in the mixed-signal after removing the stronger signals. The specific process is displayed in Fig. \ref{Hierarchical classification for mixed signals with different power}. 

The reason for considering the hierarchical structure shown in the figure above is because CNN has a higher classification and recognition accuracy for stronger signals and its time and algorithm complexity are lower. However, the recognition of weaker signals requires the use of high-precision LSTM algorithms, although the algorithms are more complex and require a high time and memory footprint.

\subsection{Design of LSTM}\label{sec:LSTM}

The deep learning model of LSTM mainly has a good classification effect for time series sequences. This part is intended to explore its classification accuracy for mixed signals under different power. Taking into account the gradient descent and gradient explosion of the LSTM model, this paper will adopt the following LSTM structure.

\begin{table}[htbp]
\renewcommand\arraystretch{2}
\caption{LSTM structure for mixed signals}
\label{table:LSTM structure for mixed signals}
\setlength{\tabcolsep}{1mm}
\begin{tabular}{|c|c|c|}
\hline
\textbf{Layer}        & \textbf{Parameter (n)} & \textbf{Activation Function} \\ \hline
Input layer           & /                  & /                            \\ \hline
LSTM layer            & 128                & /                            \\ \hline
LSTM layer            & 64                 & /                            \\ \hline
LSTM layer            & 16                 & /                            \\ \hline
Flattening layer      & /                  & /                            \\ \hline
Fully connected layer & 64                 & ReLU                         \\ \hline
Fully connected layer & 16                 & ReLU                         \\ \hline
Output layer          & 6                  & Softmax                      \\ \hline
\end{tabular}
\end{table}

In Table \ref{table:LSTM structure for mixed signals}, the parameter $n$ for LSTM represents the number of hidden layers. ReLU is used as the activation function for the fully connected layers, and Softmax is used as the activation function for the output layer.

The main design parameters of the LSTM model are the size of the hidden layers and the setting of the number of LSTM layers. After many simulations and evaluations, this model can better solve the mixed signals classification problem. It should be noticed that the data sets of LSTM and CNN will have certain differences in the input dimensions of the signal. The input data of CNN is pre-processed three-dimensional data, which can be regarded as image features for input. 
LSTM is more suitable for the processing of symbol sequences, so only the real part and imaginary part of the one-dimensional symbol sequence are preprocessed into two-dimensional symbol sequences that can meet the input requirements of the model. 

\subsection{Design of CLDNN}
The CLDNN model constructed in this paper is to connect several layers of CNN after the input layer to extract local features, and the output of CNN is poured into several layers of LSTM units to reduce time-domain changes. The output of the last layer of LSTM is input to the fully connected layer, the purpose is to map the feature space to the output layer that is easier to classify. Table \ref{table:CLDNN structure for mixed signals} shows the network structure of CLDNN.
\begin{table}[htbp]
\renewcommand\arraystretch{2}
\caption{CLDNN structure for mixed signals}
\label{table:CLDNN structure for mixed signals}
\setlength{\tabcolsep}{1mm}{
\begin{tabular}{|c|c|c|}
\hline
\textbf{Layer}        & \textbf{Parameter $(n * a * b)$} & \textbf{Activation Function} \\ \hline
Input layer           & /                  & /                            \\ \hline
Convolutional layer   & 64*2*2             & ReLU                         \\ \hline
Maximum pooling layer & 1*2                & /                            \\ \hline
Convolutional layer   & 32*1*4             & ReLU                         \\ \hline
Maximum pooling layer & 1*2                & /                            \\ \hline
Convolutional layer   & 20*1*4             & ReLU                         \\ \hline
Maximum pooling layer & 1*2                & /                            \\ \hline
Convolutional layer   & 16*1*4             & ReLU                         \\ \hline
Maximum pooling layer & 1*2                & /                            \\ \hline
Flattening layer      & /                  & /                            \\ \hline
LSTM layer            & 64                & /                            \\ \hline
LSTM layer            & 64                 & /                            \\ \hline
Fully connected layer & 256                & ReLU                         \\ \hline
Fully connected layer & 64                 & ReLU                         \\ \hline
Fully connected layer & 16                 & ReLU                         \\ \hline
Output layer          & 6                  & Softmax                      \\ \hline
\end{tabular}}
\end{table}

In Table \ref{table:CLDNN structure for mixed signals}, $n$ in the parameter $n * a * b$ represents the number of convolution kernels, and $a * b$ represents the size of the convolution kernel. The parameter $c*d$ for maximum pooling layer represents the pooling size of the pooling layer. The parameter $n$ for LSTM represents the number of hidden layers. ReLU is used as the activation function for the convolutional layer and the first three fully connected layers, and Softmax is used as the activation function for the output layer.

In addition to CLDNN, CNN, RNN and DNN also have a combination of different structures. Researchers have linearly integrated these three models. The integration model trains CNN, RNN and DNN separately, then linearly integrates the outputs of the three networks, and finally obtains the classification output. However, the performance of this model in practical applications is weaker than that of the CLDNN model.

\section{Simulation Results and Analysis\label{cha:results}}

In this section, we provide simulation and anslysis results. There are two innovations in this part: First, use the improved CNN model to identify modulated signals with different energy mixtures. By adjusting the power ratio, the number of training sets, the types of training sets and other factors, the impact on the robustness of the decision result is explored. Second, use different deep learning models for modulation and classification of different power mixed signals, and to achieve overall accuracy improvement by comparing with the CNN model used in the previous part.

We first reproduce the single-signal modulation recognition scheme based on CNN. Including the setting of training parameters and testing parameters, the training process and the theoretical analysis of the results. Next, the CNN model is partially improved, and the modulation classification of mixed signals based on CNN is realized. It includes the comparison of the recognition accuracy of mixed signals with different power ratios, and the influence of the number, type, and method of training sets on the recognition accuracy. Second, the established signal models of different powers are used as input to obtain the signal recognition accuracy of various models (ResNet, LSTM, CLDNN, and multi-level CNN/LSTM networks). According to the analysis of these results, the advantages and disadvantages of the corresponding model are obtained. Finally, the complexity of these methods will be specifically analyzed.

\subsection{Simulation and Result Analysis for CNN}
This part will use the CNN models established earlier to simulate the recognition accuracy of single signal and mixed signals, respectively. Among them, the simulation of a single signal is mainly the reproduction of the existing theory. For the simulation of mixed signals recognition accuracy, the power of the signal training set, the number of training sets, the type of training signal, and the impact of the training method will be considered. 

For the CNN training parameters, the specific selection is: the loss function is a multi-category cross-entropy loss function. The optimization method is $Adam$, the learning rate is 0.0001, the fuzzy factor is ${10}^{-8}$, the attenuation value of the learning rate after each update is 0, and the learning rate update related parameters $\beta_1 = 0.9$, $\beta_2 = 0.999$. The signal set of the training data is consistent with the test data, and the statistical model of the generalized channel experienced by the training data is consistent with the statistical model of the generalized channel experienced by the test data. For the training data, $N_{sample}=60,000$ is generated for each signal category by the Monte Carlo method, and $0.9*N_{sample}$ signals are selected for training, and the remaining $0.1*N_{sample}$ signals are selected for cross-validation. During training, the batch size is 100. After each Epoch, the internal data of the batch shuffles the selection again. The training stop condition is that the number of iterations reaches 100 Epoch or the loss value of the cross-validation set does not decrease for 10 consecutive Epochs, and the weight corresponding to the Epoch with the smallest loss value of the cross-validation set is selected as the final saved network weight.

\subsubsection{Single Signal Simulation Results and Analysis Based on CNN}
\begin{figure}[htbp]
\centering  %
\subfigure[\centering Epoch categorical accuracy for training and validation ]{
\label{Epoch categorical accuracy for training and validation}
\includegraphics[width=0.48\textwidth]{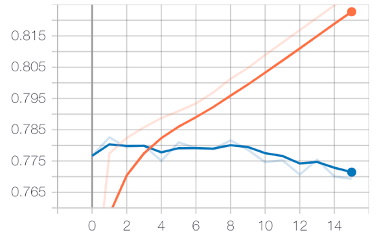}}

\subfigure[Epoch loss for training and validation]{
\label{Epoch loss for training and validation}
\includegraphics[width=0.48\textwidth]{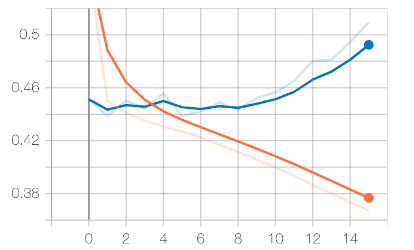}}

\caption{Training result of CNN (single signal)}
\label{Training result of CNN (single signal)}
\end{figure}

The result of the training process is shown in Fig. \ref{Training result of CNN (single signal)}.

Next are the results obtained by testing the trained model. The results include the comparison between the classification accuracy of the convolutional neural network under Gaussian white noise and the method based on the average likelihood ratio test. The classification accuracy of the lower network for each type of signal and the modulation recognition confusion matrix under different SNRs.
\begin{figure}[htbp]
\centering  
\subfigure[CNN VS. ALRT]{
\label{CNN VS. ALRT}
\includegraphics[width=0.48\textwidth]{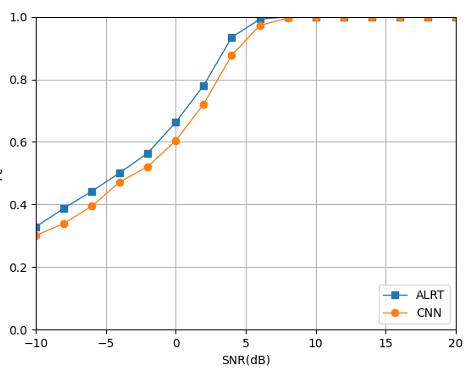}}

\subfigure[Classification accuracy of each type of signal]{
\label{Classification accuracy of each type of signal}
\includegraphics[width=0.47\textwidth]{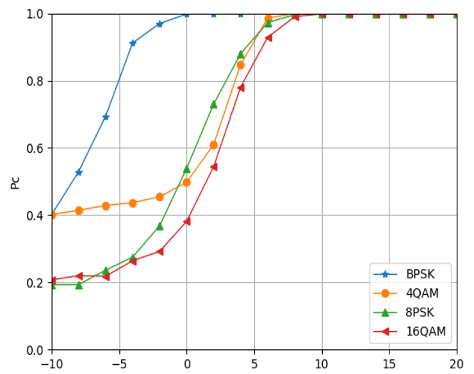}}
\caption{Classification result of CNN (single signal)-1}
\label{Classification result of CNN (single signal)-1}
\end{figure}

Fig. \ref{CNN VS. ALRT} illustrates that the CNN-based modulation classification method has a small gap from the upper bound of ALRT, indicating that the CNN method has good classification performance. Using a more complex network structure and a richer data set can further improve the classification accuracy. Fig. \ref{Classification accuracy of each type of signal} illustrates that for modulation methods with different modulation orders, low-order modulation signals require lower SNR to achieve wonderful modulation classification performance, and high-order modulation signals require higher SNR to achieve better modulation classification performance.
\begin{figure}[htbp]
	\centering
	\subfigure[Confusion matrix when SNR=-5dB]{
		\begin{minipage}[htbp]{0.4\textwidth}
			\includegraphics[width=0.95\textwidth]{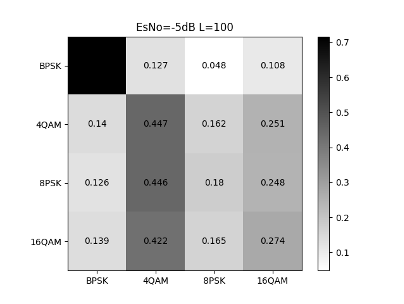}
		\end{minipage}
	}
    	\subfigure[Confusion matrix when SNR=0dB]{
    		\begin{minipage}[htbp]{0.4\textwidth}
   		 	\includegraphics[width=0.95\textwidth]{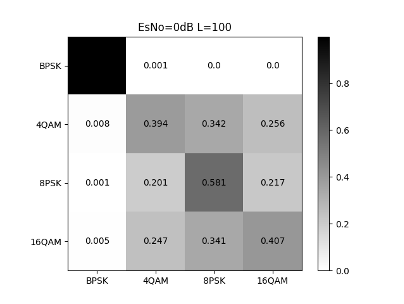}
    		\end{minipage}
    	}
	\subfigure[Confusion matrix when SNR=5dB]{
		\begin{minipage}[b]{0.4\textwidth}
			\includegraphics[width=0.95\textwidth]{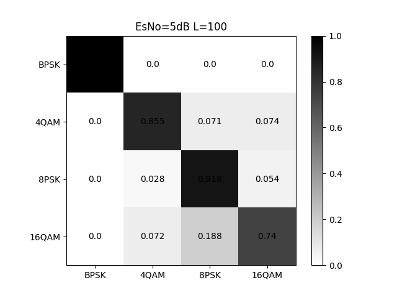}
		\end{minipage}
	}
    	\subfigure[Confusion matrix when SNR=10dB]{
    		\begin{minipage}[b]{0.4\textwidth}
		 	\includegraphics[width=0.95\textwidth]{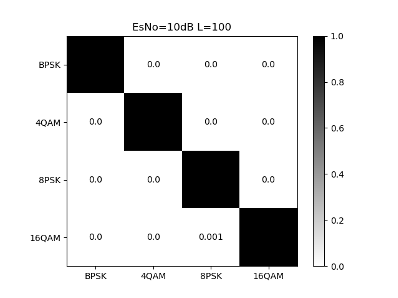}
    		\end{minipage}
    	}
	\caption{Classification result of CNN (single signal)-2}
	\label{Classification result of CNN (single signal)-2}
\end{figure}

Fig. \ref{Classification result of CNN (single signal)-2} illustrates that the CNN network's classification accuracy for a single signal is poor at low SNR. As the SNR increasing, the signal recognition rate steadily increases. When the SNR is greater than 7dB, there will be almost no misjudgment.

\subsubsection{Mixed Signals Simulation and Result Analysis Based on CNN}
\paragraph{Mixed Signals Modulation Classification Results under Same Power}
This part mainly focuses on the classification results of mixed signals under the same power. The Classification signal set is $\Omega_6=\{BPSK+4QAM,BPSK+8PSK,BPSK+16QAM, 4QAM+8PSK,4QAM+16QAM,  8PSK+16QAM\}$.$\ N_{sample}\ $ is chosen as 60000 in this part. The total power of each signal combination is normalized, which means that the signal power represented by different SNRs will be divided equally to allocate to different signals in each signal set. 
The test accuracy rate after changing the CNN structure will first be compared with that before the change. Meanwhile, the classification accuracy results of different signal combinations will also be displayed.
\begin{figure}[htbp]   
    \centering  %
    \includegraphics[width=9cm]{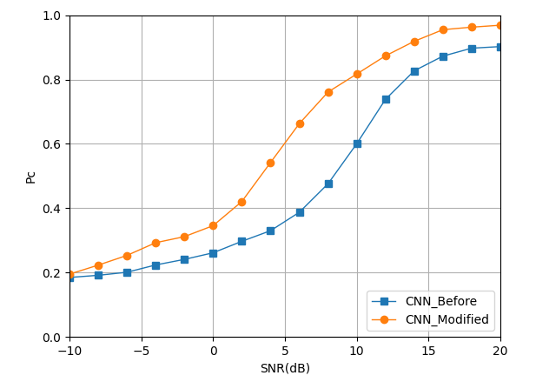}   
    \caption{Classification accuracy for CNN structure (Table \ref{table:CNN structure for single signal} Vs Table \ref{table:CNN structure for mixed signals})}    %
    \label{Classification accuracy for CNN structure (Table 3.1 Vs Table 3.2)}  
\end{figure}
\begin{figure}[htbp]   
    \centering  %
    \includegraphics[width=9cm]{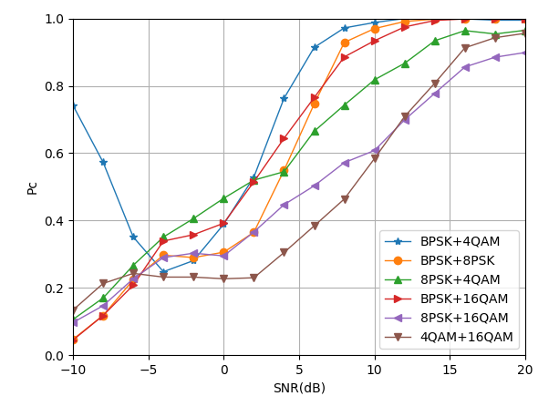}   
    \caption{Classification accuracy of each type of signals combination }    %
    \label{Classification accuracy of each type of signals combination }  %
\end{figure}

Fig. \ref{Classification accuracy for CNN structure (Table 3.1 Vs Table 3.2)} illustrates that the improved CNN structure has significantly improved the recognition accuracy of mixed signals, especially in the case of SNR=8dB, the recognition accuracy has been improved by nearly 30\%. But in the case of high SNR, the recognition accuracy does not reach 100\%. The main reason is that the combination of high-order mixed signals has a certain impact on recognition accuracy.

Through the analysis of Fig. \ref{Classification accuracy of each type of signals combination }, it can be seen that the low-order modulation signal combination requires lower SNR to achieve good modulation recognition performance, and the high-order modulation signal combination requires higher SNR to achieve good modulation recognition performance.

\paragraph{Mixed Signals Modulation Classification Results under Different Power}
One of the core issues studied in this paper is the modulation classification scheme of mixed signals of different power. Starting from this part, all the following research directions will focus on the modulation classification of mixed signals under different power. There are usually two methods for the differential representation of power in the signal set. The first type: the power of one signal is fixed, and the power of the other signal is a multiple of the current signal. The second type: limit the overall power of the mixed signals. Through the different distribution of the overall power, the different power in the modulation type combination is reflected. 

The first method will bring higher recognition accuracy because the overall signal power is larger and the influence of noise is smaller. However, this method cannot achieve the unity of the overall energy for the mixed recognition of signals of different power, and the subsequent analysis of the differential changes caused by different power cannot be done. Taking into account various factors, the second energy normalization method will be adopted for all modulation classification of mixed signals with different power in this article.

Considering the multiple combinations of signal strength and weakness, the basic signal set of this part is $\Omega_3=\{BPSK,4QAM,8PSK\}$. Taking the signal power ratio of 2:1 as an example, the total signal set type is $\Omega_6=\{2*BPSK+4QAM,2*BPSK+8PSK,2*4QAM+8PSK,2*4QAM+BPSK, 2*8PSK+BPSK,2*8PSK+4QAM\}$. $N_{sample}\ $ is chosen as 60000 in this part and the CNN structure used is shown in Table \ref{table:CNN structure for mixed signals}. This part first compares the accuracy of mixed signals recognition with a power ratio of 2:1 with the upper limit of ALRT. Next, compare the recognition accuracy of mixed signals under different power ratios (2:1, 5:1, 8:1). Finally, signals with different power ratios (1:1 to 9:1) are randomly generated as the training set to realize the recognition and classification of random mixed power signals by the CNN model.
\begin{figure}[htbp]   
    \centering  
    \includegraphics[width=8.5cm]{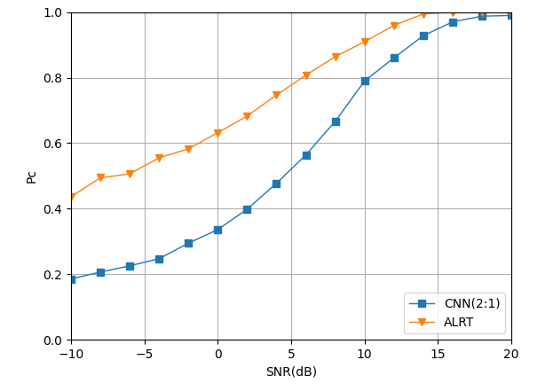}   
    \caption{CNN compared with ALRT (signal power ratio: 2:1)}   
    \label{CNN compared with ALRT (signal power ratio: 2:1) }  
\end{figure}

Fig. \ref{CNN compared with ALRT (signal power ratio: 2:1) }   shows the classification results of the CNN structure shown in Table \ref{table:CNN structure for mixed signals} for a mixed-signal with a power ratio of 2:1. It can be clearly seen from the figure that SNR has an impact on recognition accuracy. Besides, compared with the upper limit of recognition accuracy (ALRT), the recognition accuracy under low SNR still has a large room for improvement, and the recognition accuracy under high SNR has a smaller gap compared with the theoretical upper limit.
\begin{figure}[htbp]   
    \centering  
    \includegraphics[width=9cm]{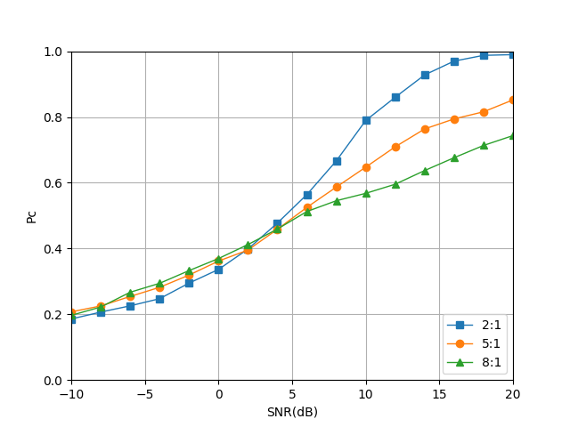}   
   \caption{Classification accuracy for CNN under signal power ratio: 2:1; 5:1; 8:1}
     \label{Classification accuracy for CNN under signal power ratio: 2:1; 5:1; 8:1 }  
\end{figure}

Fig. \ref{Classification accuracy for CNN under signal power ratio: 2:1; 5:1; 8:1 }  shows the classification consequences of mixed signals with different power ratios. It can be clearly seen from the figure that in the SNR range below 5 dB, the mixing of signals of different power will not have much impact on the accuracy of recognition. However, as the SNR increases, signals with smaller power ratios have higher recognition accuracy. The main reason is that when the ratio of signal power differs greatly, the signal with less power is treated as noise, which results in a misjudgment of the classification result of the signal combination.
\begin{figure}[htbp]   %
    \centering  %
    \includegraphics[width=9cm]{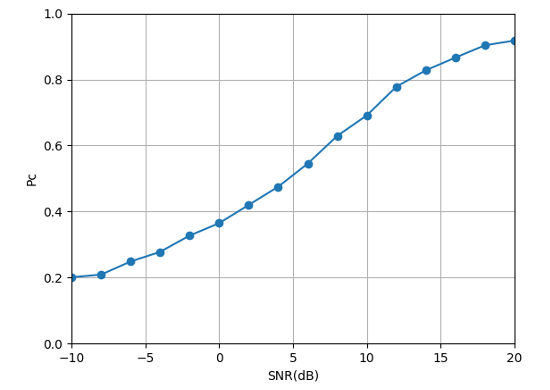}   %
    \caption{Classification accuracy for mixed signals with different signal power ratios (1:1 to 9:1)}
    \label{Classification accuracy for mixed signals with different signal power ratios (1:1 to 9:1)1 }  %
\end{figure}
\begin{figure}[htbp]   %
    \centering  %
    \includegraphics[width=9cm]{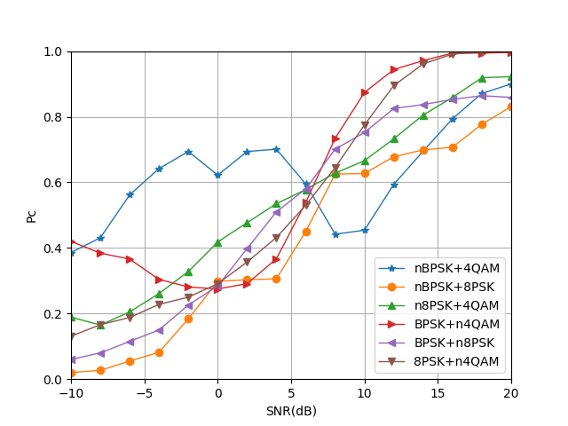}   %
    \caption{Classification accuracy of each type of signals combination (random signal power ratio)}
    \label{Classification accuracy of each type of signals combination (random signal power ratio) }  %
\end{figure}
\begin{figure}
	\centering
	\subfigure[Confusion matrix when SNR=-10dB]{
		\begin{minipage}[htbp]{0.35\textwidth}
			\includegraphics[width=0.9\textwidth]{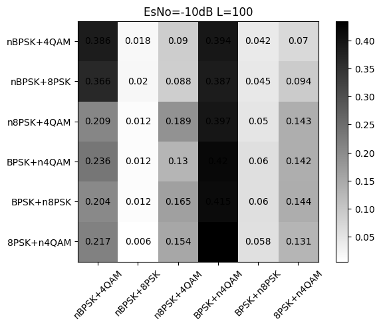}
		\end{minipage}
	}
    	\subfigure[Confusion matrix when SNR=0dB]{
    		\begin{minipage}[htbp]{0.35\textwidth}
   		 	\includegraphics[width=0.9\textwidth]{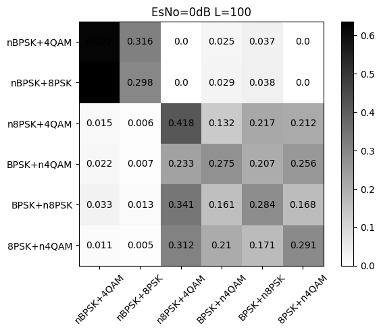}
    		\end{minipage}
    	}
	\subfigure[Confusion matrix when SNR=10dB]{
		\begin{minipage}[b]{0.35\textwidth}
			\includegraphics[width=0.9\textwidth]{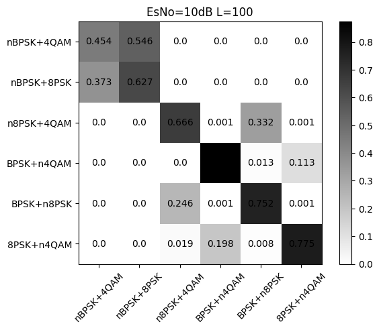}
		\end{minipage}
	}
    	\subfigure[Confusion matrix when SNR=20dB]{
    		\begin{minipage}[b]{0.35\textwidth}
		 	\includegraphics[width=0.9\textwidth]{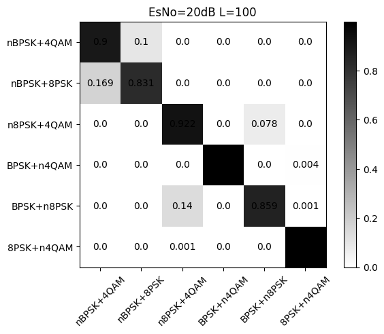}
    		\end{minipage}
    	}
	\caption{Classification accuracy for mixed signals with different signal power ratios (1:1 to 9:1)}
	\label{Classification accuracy for mixed signals with different signal power ratios (1:1 to 9:1)2}
\end{figure}

Fig.  \ref{Classification accuracy for mixed signals with different signal power ratios (1:1 to 9:1)1 } to Figure \ref{Classification accuracy for mixed signals with different signal power ratios (1:1 to 9:1)2} show the recognition accuracy of the basic CNN structure for modulated signals with random mixed power ratios. It can be seen that there is a certain recognition accuracy (92\%) under high SNR, but the recognition accuracy is low under low SNR. From Fig. \ref{Classification accuracy of each type of signals combination (random signal power ratio) } and Fig. \ref{Classification accuracy for mixed signals with different signal power ratios (1:1 to 9:1)2}, we can see: First, the high recognition accuracy after high-order signal mixing needs to be achieved under higher SNR, so the overall recognition accuracy influences. Second, the signal with lower power in the mixed-signal will be misjudged as noise. Although the signal with high power can be identified, a deviation occurs when distinguishing the mixed signals. Taking n$*$BPSK+4QAM and n$*$BPSK+8PSK as examples, when the SNR is 20dB, the error probability between these two parts is close to 0.15.

\begin{figure}[htbp]   %
    \centering  %
    \includegraphics[width=7.5cm]{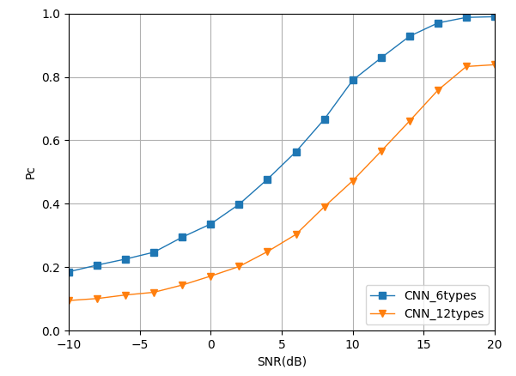}   %
    \caption{Comparison of CNN classification accuracy before and after the signal category is increased} 
    \label{Comparison of CNN classification accuracy before and after the signal category is increased }  %
\end{figure}

\begin{figure}[htbp]   %
    \centering  %
    \includegraphics[width=8cm]{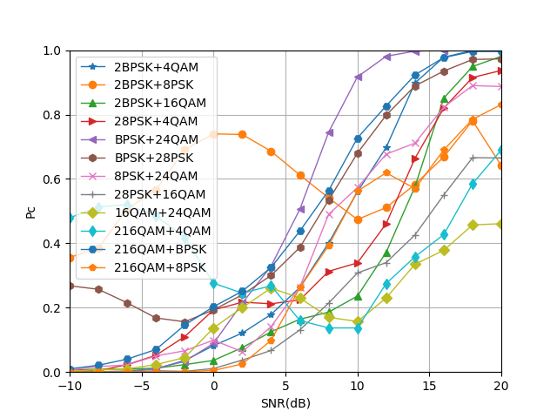}   %
    \caption{Classification accuracy of CNN after the signal category is increased}    %
    \label{Classification accuracy of CNN after the signal category is increased}  %
\end{figure}

\begin{figure}
	\centering
	\subfigure[Confusion matrix when SNR=5dB]{
		\begin{minipage}[htbp]{0.33\textwidth}
			\includegraphics[width=0.9\textwidth]{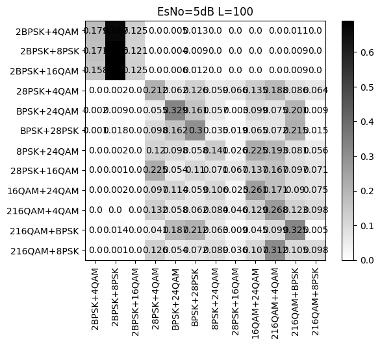}
		\end{minipage}
	}
    	\subfigure[Confusion matrix when SNR=10dB]{
    		\begin{minipage}[htbp]{0.33\textwidth}
   		 	\includegraphics[width=0.9\textwidth]{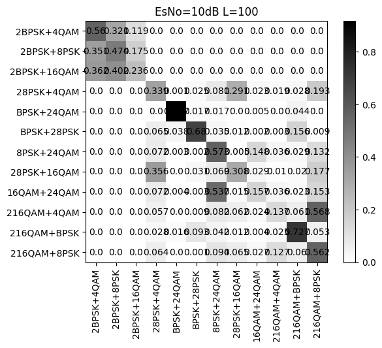}
    		\end{minipage}
    	}
	\subfigure[Confusion matrix when SNR=15dB]{
		\begin{minipage}[b]{0.33\textwidth}
			\includegraphics[width=0.9\textwidth]{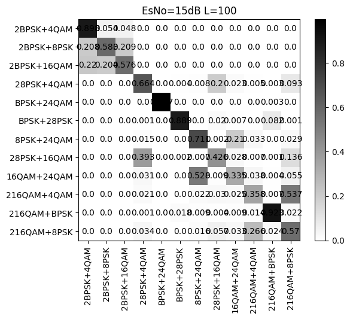}
		\end{minipage}
	}
    	\subfigure[Confusion matrix when SNR=20dB]{
    		\begin{minipage}[b]{0.33\textwidth}
		 	\includegraphics[width=0.9\textwidth]{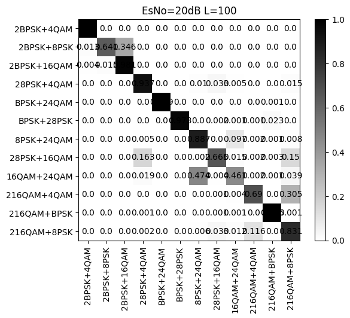}
    		\end{minipage}
    	}
	\caption{CNN's modulation classification confusion matrix after the signal category is increased }
	\label{CNN's modulation classification confusion matrix after the signal category is increased }
\end{figure}

\paragraph{The Impact of Number of Signal Categories on Classification Accuracy}
To explore the impact of the improvement of mixed-signal types with different power on the recognition accuracy of CNN. This section sets the basic signal set of the signal as $\Omega_4=\{BPSK,4QAM,8PSK,16QAM\}$, the signal power ratio is set as 2:1, so the total signal set type is $\Omega_{12}=\{ 2*BPSK+4QAM, 2*BPSK+8PSK,2*BPSK+16QAM, 2*4QAM+BPSK, 2*4QAM+8PSK, 2*4QAM +16QAM, 2*8PSK+BPSK, 2*8PSK+ 4QAM,2*8PSK+16QAM,2*16QAM+BPSK,2*16QAM+4QAM, 2*16QAM+8PSK\}$.  $N_{sample}$  is chosen as 60000 in this part and the CNN structure used is shown in Table \ref{table:CNN structure for mixed signals}. Other settings are consistent with the previous descriptions in this section.
Next, compare the classification accuracy of the two-by-two mixture of four different signals with different power and that of the three-signal mixture, and the results are shown in Figs. \ref{Comparison of CNN classification accuracy before and after the signal category is increased }, \ref{Classification accuracy of CNN after the signal category is increased} and \ref{CNN's modulation classification confusion matrix after the signal category is increased }.

Fig. \ref{Comparison of CNN classification accuracy before and after the signal category is increased }  illustrates that the increase in the types of mixed signals has a significant impact on the classification accuracy of CNN, and the overall lag is nearly 7 dB.

Fig. \ref{Classification accuracy of CNN after the signal category is increased} illustrates that the combination of high-order signals still produces misjudgment in the high SNR area, especially for the $16QAM+2*4QAM$ combination mode. When the SNR is 20dB, the recognition accuracy is only close to 50\%. The confusion matrix shown in Fig. \ref{CNN's modulation classification confusion matrix after the signal category is increased } can better illustrate this point. The main reasons for misjudgment are concentrated in the mixed signals containing QAM signals which illustrates that the mixed signals containing QAM signals will have a certain impact on the recognition accuracy of CNN.

\paragraph{The Impact of Number of Training Sets on Classification Accuracy}
The training set size refers to the number of symbol sequences used for training. This section uses simulation experiments to observe the modulation recognition performance under different training set settings to reflect the impact of the training set size on the classification accuracy. The signal power ratio is set as 2:1 and the total signal set type is $\Omega_6=\{2*BPSK+4QAM,2*BPSK+8PSK,2*4QAM+8PSK,2*4QAM+BPSK,2*8PSK+BPSK,2*8PSK+4QAM\}$. The CNN structure used is shown in Table \ref{table:CNN structure for mixed signals} and $N_{sample}\ $ is chosen as 5000, 10000, 60000, 120000 respectively in this part.
\begin{figure}[htbp]   %
    \centering  %
    \includegraphics[width=9cm]{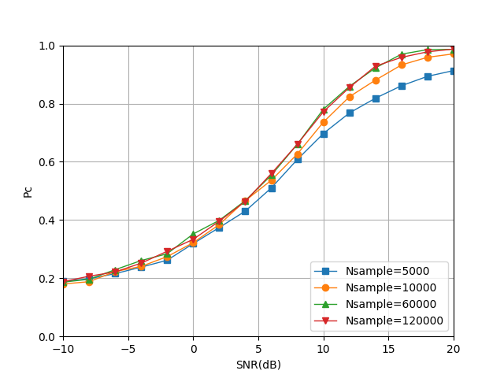}   %
    \caption{Impact of the number of training samples on the accuracy of CNN recognition}
    \label{Impact of the number of training samples on the accuracy of CNN recognition}  %
\end{figure}

The simulation results are shown in Fig.  \ref{Impact of the number of training samples on the accuracy of CNN recognition}. It can be seen that the larger the amount of training data used, the better the modulation recognition performance of the trained CNN. When the quantity of training data is small, increasing the quantity of training data can significantly improve the modulation recognition performance of the trained CNN. When the quantity of training data increases to a certain extent, the effect of improving the modulation recognition performance of the trained CNN will be reduced. Therefore, from the perspective of performance guarantee and complexity, the number of training samples is designed to $N_{sample}$ = 60000.

\paragraph{The Impact of Training Set under Fixed SNR on Classification Accuracy}
The previous setting of the training set is to randomly generate signal combinations of different modulation types under different SNRs. To explore the impact of the training set under a fixed SNR on recognition accuracy. In this part, the corresponding training set will be established when SNR=-10dB, -5dB, 0dB, 5dB, 10dB, 15dB, 20dB, and the overall recognition accuracy curve will be described. The signal power ratio is set as 2:1 and the total signal set type is  $\Omega_6=\{2*BPSK+4QAM,2*BPSK+8PSK,2*4QAM+8PSK,2*4QAM+BPSK,\ 2*8PSK+BPSK,2*8PSK+4QAM\}$.  The CNN structure used is shown in Table \ref{table:CNN structure for mixed signals} and $N_{sample}\ $ is chosen as 60000.
\begin{figure}[htbp]   %
    \centering  %
    \includegraphics[width=9cm]{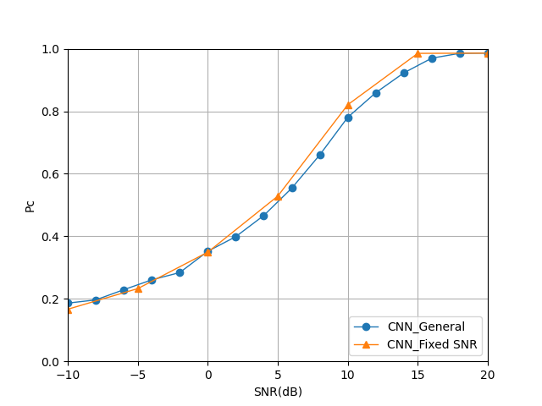}   %
    \caption{ Effect of fixed SNR training samples on CNN recognition accuracy}    %
    \label{ Effect of fixed SNR training samples on CNN recognition accuracy}  %
\end{figure}

Fig. \ref{ Effect of fixed SNR training samples on CNN recognition accuracy}  illustrates the fixed SNR training mode has a certain improvement in the recognition accuracy of CNN. But the range of improvement is mainly concentrated in the area after 0dB, which does not play a big role in the recognition rate under low SNR.

\subsection{Simulation and Result Analysis for Five Deep Learning Models}
This part mainly uses the various deep learning models mentioned earlier to simulate the results of the combined classification of mixed signals under different power ratios. The basic signal set of this part is $\Omega_3=\left\{BPSK,4QAM,8PSK\right.\}$, 
so the total combination set $\Omega_6=\{n*BPSK+4QAM,n*BPSK+8PSK,n*4QAM+BPSK,n*4QAM+8PSK,n*8PSK+BPSK, n*8PSK+4QAM.\}$. 
 Different power ratio (2:1, 5:1, 8:1) will be considered in this part which means n= 2,5,8. Other parameters will still be the same, signal sequence length $L = 100$, the $i_{th}$ received signal sequence satisfies $r_i(n)=s_i(n)+g_i(n)$, where $n\in\{1,\cdots,L\}, g_i(n) \sim CN(0,1)$. 
 
 The average SNR of each symbol sequence obeys a uniform distribution between [-10dB, 20dB] and the noise is different. The training data adopts a Monte Carlo method to generate $N_{sample}=60000$ symbol sequences for each modulation category, and a total of 
 $6{\ast N}_{sample}=360000$ 
signal sequences are generated. Among them, $0.1*N_{sample}=6000$ data is used as the cross-validation set, and the rest is used as the training set. The test data uses the same method. The SNR test range is [-10dB, 20dB], and the test interval is 2dB. The symbol sequence of each modulation category corresponds to a specific SNR to generate 1000 test sets. Finally, the test results are averaged to obtain the modulation recognition accuracy rate.

Besides, the loss function is a multi-category cross-entropy loss function. The optimization method is Adam, the learning rate is 0.0001, the fuzzy factor is ${10}^{-8}$, the attenuation value of the learning rate after each update is 0, and the learning rate update related parameters $\beta_1 = 0.9, \beta_2 = 0.999$. The signal set of the training data is consistent with the test data, and the statistical model of the generalized channel experienced by the training data is consistent with the statistical model of the generalized channel experienced by the test data. During training, the batch size is 100. After each Epoch, the internal data of the batch shuffles the selection again. The training stop condition is that the number of iterations reaches 100 Epoch or the loss value of the cross-validation set does not decrease for 10 consecutive Epochs, and the weight corresponding to the Epoch with the smallest loss value of the cross-validation set is selected as the final saved network weight.

\subsection{Comparison of Classification Performance under Signal Power Ratio of 2:1}
When the power ratio is 2:1, the basic combination set $\Omega_6=\{2*BPSK+4QAM,2*BPSK+8PSK,2*4QAM+8PSK,2*4QAM+BPSK,2*8PSK+BPSK, 2*8PSK+4QAM\}$.  The results should be shown below.
\begin{figure}[htbp]   %
    \centering  %
    \includegraphics[width=9cm]{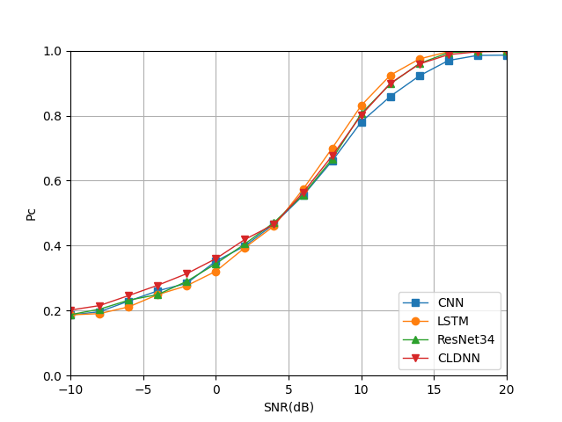}   %
    \caption{Classification accuracy under signal power ratio of 2:1}
    \label{Classification accuracy under signal power ratio of 2:1}  %
\end{figure}

From Fig. \ref{Classification accuracy under signal power ratio of 2:1}, it can be seen that when the mixed signals power ratio is 2:1, when the SNR is greater than 10dB, the classification accuracy of the other three deep learning models for mixed signals is greater than the CNN designed in Section III, and LSTM has the highest recognition accuracy. But in the low SNR area (SNR<0dB), the recognition accuracy of LSTM is low, and CLDNN has higher recognition accuracy than other deep learning models. On the whole, CLDNN has a better overall performance.

\subsubsection{Comparison of Classification Performance under Signal Power Ratio of 5:1}
When the power ratio is 5:1, the basic combination set $\Omega_6=\{5*BPSK+4QAM,5*BPSK+8PSK,5*4QAM+8PSK,5*4QAM+BPSK, 5*8PSK+BPSK, 5*8PSK+4QAM\}$.  The results should be shown below.

From Fig. \ref{Classification accuracy under signal power ratio of 5:1}, we can see when the SNR is low (less than 5dB), the classification and recognition accuracy of the four deep learning models for mixed signals are relatively similar. However, when the SNR gradually becomes larger, the LSTM and ResNet34 models show better recognition accuracy, and even with the continued increase of SNR, there is a tendency to approach 100\%. However, these two deep learning models are more complex and require more time and memory to train the training set. It can be said that the algorithm complexity is sacrificed in exchange for higher recognition efficiency.
\begin{figure}[htbp]   %
    \centering  %
    \includegraphics[width=9cm]{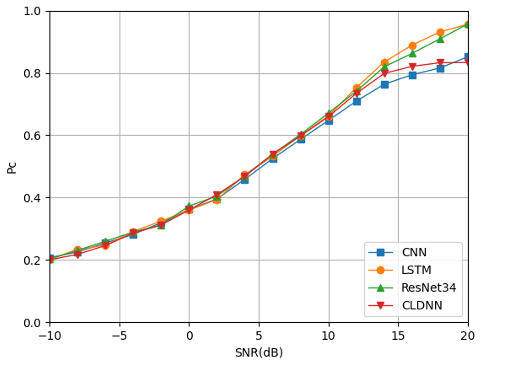}   %
    \caption{Classification accuracy under signal power ratio of 5:1}
    \label{Classification accuracy under signal power ratio of 5:1}  %
\end{figure}

\subsubsection{Comparison of Classification Performance under Signal Power Ratio of 8:1}
\begin{figure}[htbp]  
    \centering  %
    \includegraphics[width=9cm]{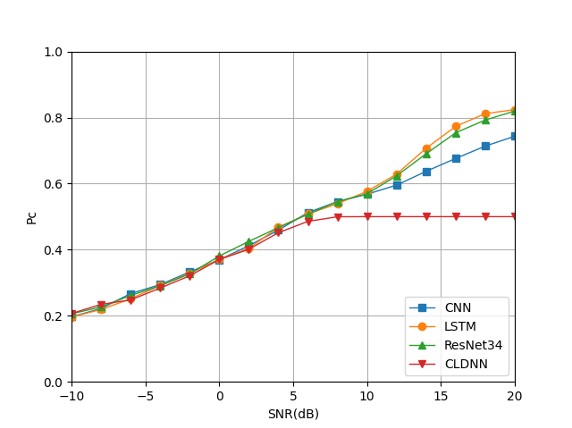}   %
    \caption{Classification accuracy under signal power ratio of 8:1}
    \label{Classification accuracy under signal power ratio of 8:1}  %
\end{figure}

When the power ratio is 8:1, the basic combination set $\Omega_6=\{8*BPSK+4QAM,8*BPSK+8PSK,8*4QAM+8PSK,8*4QAM+BPSK, 8*8PSK+BPSK, 8*8PSK+4QAM\}$.  The results are shown in Fig.  \ref{Classification accuracy under signal power ratio of 8:1}.
It can be seen that as the signal power strength ratio increases, the four types of deep learning models all misclassify the weaker signals in the mixed-signal, resulting in a decrease in recognition accuracy. Compared with the traditional CNN structure, the LSTM model and the ResNet model have higher recognition accuracy in the range of SNR>10dB, while the performance of CLDNN is more general. In the low SNR area, the four models failed to obtain better recognition accuracy.

\subsubsection{Comparison of Classification Performance under Signal Power Ratio of n:1}
\begin{figure}[htbp]   %
    \centering  %
    \includegraphics[width=9cm]{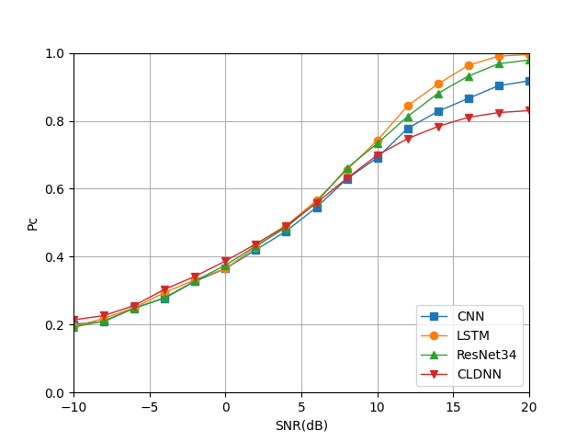}   %
    \caption{Classification accuracy under signal power ratio of n:1 (n=1,$\cdots$,9)}
    \label{Classification accuracy under signal power ratio of n:1 (n=1-9)}  %
\end{figure}

Signals with different power ratios (1:1 to 9:1) are randomly generated as the training set to realize the recognition and classification of random mixed power signals which means the basic combination set $\Omega_6=\{n*BPSK+4QAM,n*BPSK+8PSK,n*4QAM+8PSK,n*4QAM+BPSK, n*8PSK+BPSK, n*8PSK+4QAM\}$.  The results are shown in Fig. \ref{Classification accuracy under signal power ratio of n:1 (n=1-9)}.

When the power ratio of the mixed signals is randomly generated between 1:1 and 1:9 as the training set, Fig. \ref{Classification accuracy under signal power ratio of n:1 (n=1-9)} shows that various deep learning models have different accuracy for mixed-signal classification under different SNRs. In the case of low SNR (SNR $<$ 5dB), CLDNN has a slight advantage in classification accuracy compared to other deep learning models. The main reason is that it combines the feature extraction of CNN and the cyclic accumulation algorithm of LSTM, which leads to the ability of this algorithm to identify small features in the symbol sequence. When the SNR is greater than 5dB, the order of recognition accuracy is: LSTM $>$ ResNet $>$ CNN $>$ CLDNN. In general, LSTM has better performance.

\subsubsection{Hierarchical CNN's Mixed Signals Recognition Capabilities}
For the structural framework of hierarchical deep learning methods, the CNN used is the CNN structure shown in Table \ref{table:CNN structure for mixed signals}, the LSTM model is the one shown in Section III, and the training set used is a combination of training sets with a mixed signals power ratio of 2:1, $\Omega_6=\{2*BPSK+4QAM,2*BPSK+8PSK,2*4QAM+8PSK,2*4QAM+BPSK, 2*8PSK+BPSK, 2*8PSK+4QAM\}$.

The first level of classification is mainly to identify the stronger signals in the signal combination, so the influence of the weaker signals can be ignored, and the collection of stronger signals is regarded as labels for training. Therefore, the previous signal set can be divided into: $\Omega_3=\{ \{2*BPSK+4QAM,2*BPSK+8PSK\},\{2*4QAM+BPSK,2*4QAM+8PSK\},\{2*8PSK+BPSK,2*8PSK+4QAM\}\} $. The training and test results are shown in the figures below.
\begin{figure}[htbp]   %
    \centering  %
    \includegraphics[width=8cm]{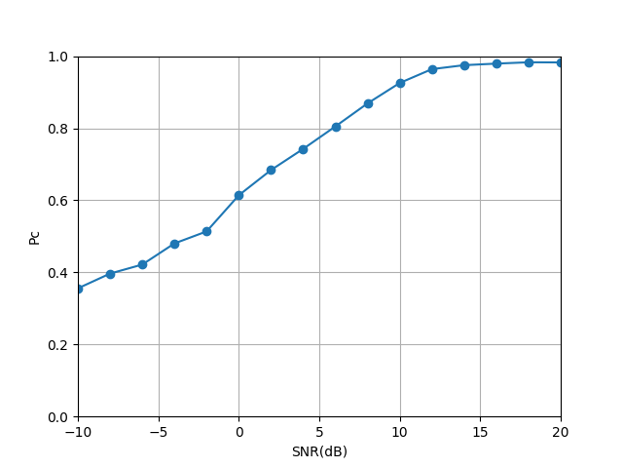}   %
    \caption [Classification accuracy of the first level]{\centering Classification accuracy of the first level}   
    \label{Classification accuracy of the first level}  %
\end{figure}
\begin{figure}[htbp]   %
    \centering  %
    \includegraphics[width=8cm]{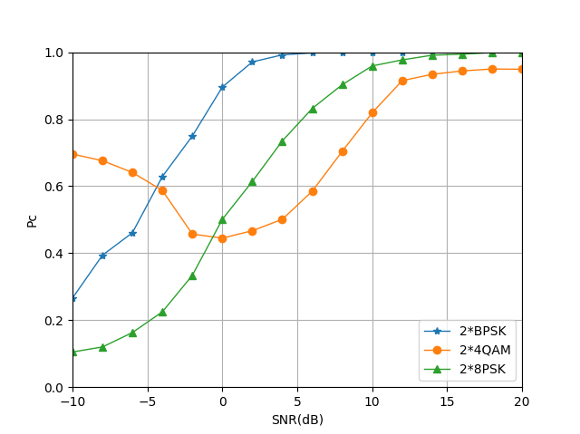}   %
    \caption [Classification accuracy of various strong signals at the first level]{\centering Classification accuracy of various strong signals at the first level}    %
    \label{Classification accuracy of various strong signals at the first level}  %
\end{figure}
\begin{figure}
	\centering
	\subfigure[Confusion matrix when SNR=5dB]{
		\begin{minipage}[htbp]{0.4\textwidth}
			\includegraphics[width=0.95\textwidth]{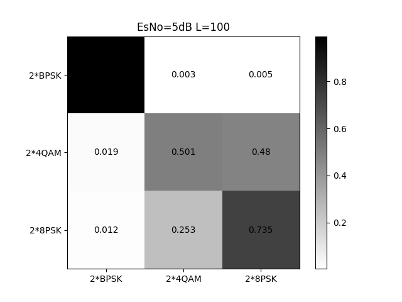}
		\end{minipage}
	}
    	\subfigure[Confusion matrix when SNR=10dB]{
    		\begin{minipage}[htbp]{0.4\textwidth}
   		 	\includegraphics[width=0.95\textwidth]{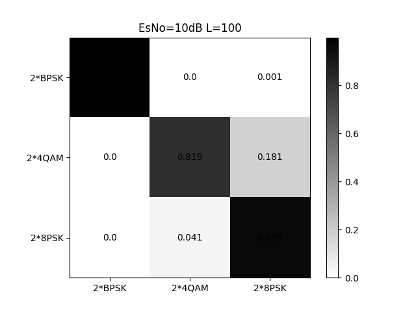}
    		\end{minipage}
    	}
	\subfigure[Confusion matrix when SNR=15dB]{
		\begin{minipage}[b]{0.4\textwidth}
			\includegraphics[width=0.95\textwidth]{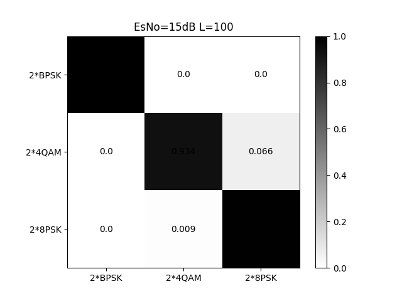}
		\end{minipage}
	}
    	\subfigure[Confusion matrix when SNR=20dB]{
    		\begin{minipage}[b]{0.4\textwidth}
		 	\includegraphics[width=0.95\textwidth]{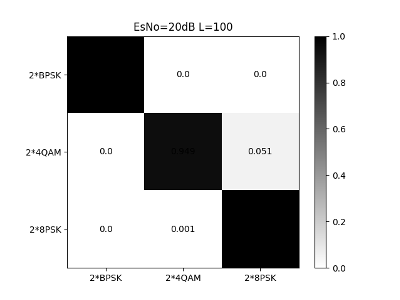}
    		\end{minipage}
    	}
	\caption{Recognition confusion matrix after the first level classification }
	\label{Recognition confusion matrix after the first level classification }
\end{figure}

From Figs. \ref{Classification accuracy of the first level}, \ref{Classification accuracy of various strong signals at the first level} and \ref{Recognition confusion matrix after the first level classification }, we can see that the CNN structure shown in Table \ref{table:CNN structure for mixed signals} has a high recognition rate for the stronger signals in the mixed signals. Except that 4QAM has a 5\% chance of being misjudged as 8PSK in high SNR areas, the overall recognition accuracy can reach 98\% at 12dB.

After distinguishing the strong signal, the LSTM structure shown in Section \ref{sec:LSTM} is used to realize the identification of the weaker signal in each category. The specific results are shown below.
\begin{figure}[htbp]   %
    \centering  %
    \includegraphics[width=9cm]{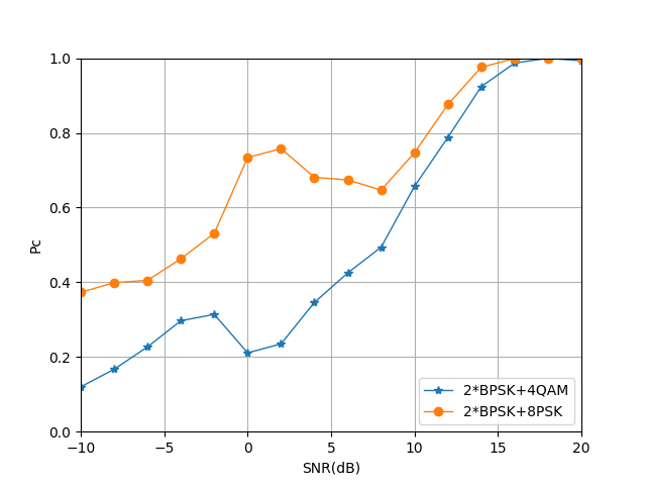}   %
    \caption{Classification result of 2$*$BPSK+4QAM,2$*$BPSK+8PSK}
    \label{Classification result of 2$*$BPSK+4QAM,2$*$BPSK+8PSK}  %
\end{figure}
\begin{figure}[htb]   %
    \centering  %
    \includegraphics[width=9cm]{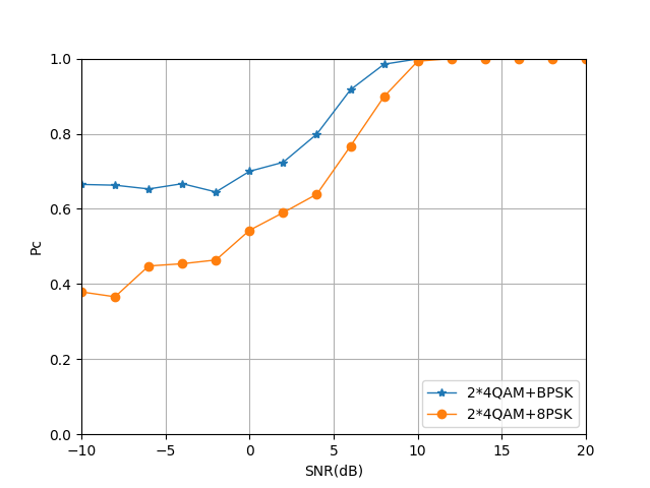}   %
    \caption{Classification result of 2$*$4QAM+BPSK,2$*$4QAM+8PSK}
    \label{Classification result of 2$*$4QAM+BPSK,2$*$4QAM+8PSK}  %
\end{figure}
\begin{figure}[htb]   %
    \centering  %
    \includegraphics[width=9cm]{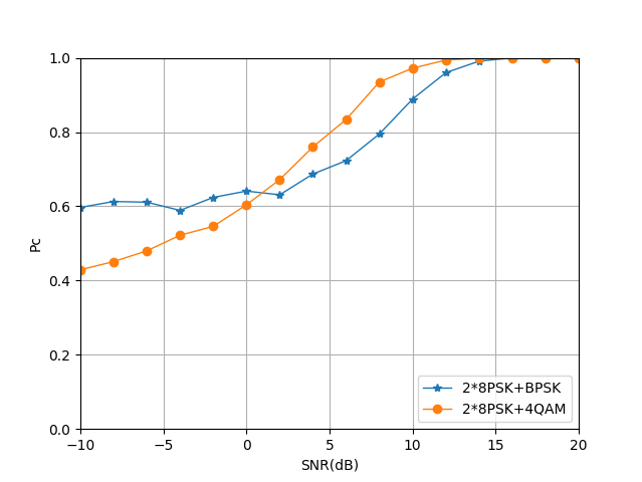}   %
    \caption{Classification result of 2$*$8PSK+BPSK,2$*$8PSK+4QAM}
    \label{Classification result of 2$*$8PSK+BPSK,2$*$8PSK+4QAM}  %
\end{figure}

From Figs. \ref{Classification result of 2$*$BPSK+4QAM,2$*$BPSK+8PSK}, \ref{Classification result of 2$*$4QAM+BPSK,2$*$4QAM+8PSK} and \ref{Classification result of 2$*$8PSK+BPSK,2$*$8PSK+4QAM}, we can see that after distinguishing the stronger signal in the mixed-signal, the identification of the weaker signal can be realized by the LSTM algorithm. On the whole, the classification accuracy of each signal can reach 100\% in the area of SNR>15dB. The mixed-signal based on 2$*$4QAM and 2$*$8PSK can also obtain better signal recognition accuracy in the low SNR area, while the recognition accuracy of the mixed signal is based on 2$*$BPSK in the low SNR is not very ideal. Based on the results of the first-level classification and the second-level classification, the classification accuracy data of mixed signals of different power ratios can be obtained as shown in table \ref{table:Comprehensive classification accuracy with an power ratio of 2:1}.
\begin{table}[htbp]
\renewcommand\arraystretch{2}
\caption{Comprehensive classification accuracy with an power ratio of 2:1}
\label{table:Comprehensive classification accuracy with an power ratio of 2:1}
\setlength{\tabcolsep}{1mm}{
\begin{tabular}{|c|c|c|c|c|c|c|c|}
\hline
\textbf{SNR   (dB)} & -10   & -5    & 0     & 5     & 10    & 15    & 20    \\ \hline
\textbf{Accuracy}   & 17.2\% & 25.6\% & 38.1\%      & 61.2\%      & 83.4\%       & 97.7\%       & 98.9\%       \\ \hline
\end{tabular}}
\end{table}

Draw the corresponding curve according to the calculated result and compare it with the result obtained before under the same conditions. It can be seen that the accuracy of the mixed signals recognition by this method is significantly improved in the range of 0-10dB.
\begin{figure}[htbp]   %
    \centering  %
    \includegraphics[width=8cm]{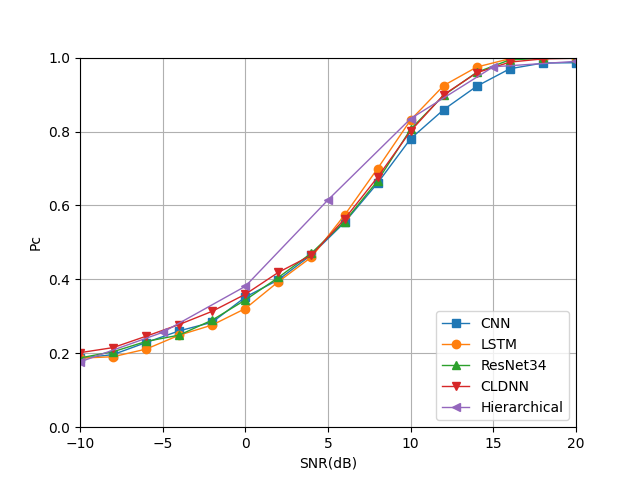}   %
    \caption{Classification accuracy under signal power ratio of 2:1 mixed}
    \label{Classification accuracy under signal power ratio of 2:1 mixed}  %
\end{figure}

\subsubsection{Algorithm Complexity Analysis}
After the analysis of the algorithm recognition accuracy problem is realized, there are two main indicators for further evaluation of the algorithm: 1) The computing power required for forwarding propagation, which reflects the level of performance requirements for hardware such as GPU; 2) Parameters Number, it reflects the size of memory occupied. The number of parameters can be obtained directly when using TensorFlow for model training, and the computing power required for forwarding propagation is reflected by FLOPs. This part will mainly calculate and analyze the FLOPs and the number of parameters of the five deep learning models mentioned above.
\begin{table}[htbp]
\renewcommand\arraystretch{2}
\caption{FLOPs and Parameters for Deep learning models }
\label{table:FLOPs and Parameters for Deep learning models }
\setlength{\tabcolsep}{1mm}{
\begin{tabular}{|c|c|c|c|}
\hline
\textbf{Model   Name} & \textbf{Model   Size (Params)} & \textbf{Model   Size (MB)} & \textbf{MFLOPs} \\ \hline
CNN   (Table \ref{table:CNN structure for mixed signals})                  & 42554    & 0.16  & 1.0   \\ \hline
ResNet34                           & 22683270 & 86.50 & 254.5 \\ \hline
LSTM                               & 231670   & 0.88  & 2.4   \\ \hline
CLDNN                              & 1915194  & 7.30  & 38.6  \\ \hline
\begin{tabular}[c]{@{}c@{}}Hierarchical \\ Structure (1st layer)\end{tabular} & 42554    & 0.16  & 1.0   \\ \hline
\begin{tabular}[c]{@{}c@{}}Hierarchical \\ Structure (2nd layer)\end{tabular} & 231670   & 0.88  & 2.4   \\ \hline
\end{tabular}}
\end{table}

According to the complexity results of the algorithm model obtained in Table \ref{table:FLOPs and Parameters for Deep learning models }, it can be found that the GPU performance required by the basic CNN is lower and the memory occupied is small, and ResNet34 needs to occupy higher memory and higher floating-point computing efficiency. Although the hierarchical structure has a high recognition accuracy for mixed signals in the middle SNR interval, this structure not only increases the complexity of the algorithm structure but also increases a certain amount of memory consumption. Combining the results of recognition accuracy and algorithm complexity, it illustrates that the LSTM model has better performance.

\section{Conclusion\label{cha:conclusion}}

This paper studied the mixed signals modulation classification technology based on deep learning. We first reviewed the existing signal modulation classification technology, and the research based on deep learning. Then we analyzed and simulated the modulation classification of the single signal based on the existing CNN model. Considering that NOMA technology has deeper requirements for the modulation classification of mixed signals under different power, this paper improved the basic CNN and applied it to the modulation classification of mixed signals. Simultaneously, the effect of the quantity of training sets, the type of training set and the training method on the accuracy of mixed signals recognition was studied. Then, the basic CNN structure was improved, and other deep learning models such as ResNet34, LSTM, hierarchical structure and CLDNN were used to further study the mixed signals modulation classification problem. This paper mainly completed the following two aspects of work:

First, consider CNN as a feature extractor and classifier simultaneously to study the CNN model whose input is a sequence of baseband symbols and output is the decision result of the modulation system. Improve the design of CNN in single-signal and mixed-signal modes, and explore the influence of different training set modes on recognition accuracy. Theoretical analysis and simulation results show that the modulation classification technology of a single signal based on CNN is basically close to the theoretical upper limit, but the modulation classification accuracy of mixed signals under different energies is still far from the upper limit. The difference in the power ratio of the mixed signals, the size of the training set, and the type of training set all have an impact on the classification accuracy. But overall, the trained CNN model can maintain good modulation classification performance for mixed signals.

Second, considering that the main content of the CNN part is to explore the impact of changes in the training set on recognition accuracy without changing the deep learning model. The following content is mainly to explore the performance of different deep learning models for the recognition accuracy of mixed signals under different powers under the condition of fixing the corresponding number of training sets and types of training sets. The four deep learning models used are ResNet34, Hierarchical structure, LSTM and CLDNN. Through theoretical analysis and simulation results, although the time and algorithm complexity have increased, compared with the CNN model, the recognition accuracy of these four deep learning models has increased in different SNR ranges. Among them, it can be found that when the power difference is small, the CLDNN model has a certain improvement in recognition accuracy in low SNR regions, and the hierarchical structure has outstanding recognition ability in medium SNR regions, LSTM model has higher recognition accuracy in high SNR areas. When the power difference is large, both LSTM and ResNet models have a good classification and recognition accuracy, while CLDNN performs poorly. In short, different deep learning models have different effects on the modulation classification problem of mixed signals at different energies.


 So far, the mixed-signal modulation classification problem based on deep learning is still in its infancy, and most scholars only stay at the problem of single-signal modulation classification. This paper only does a preliminary study on this problem, there are still a lot of problems to be solved. The research directions in the future are as follows:

 1. This paper only studies the mixed signals modulation classification based on some deep learning algorithms, and there is still a certain gap from the theoretical upper limit. New deep learning frameworks such as capsule networks and those models developed in our existing works, e.g., \cite{IREALCARE1,IREALCARE2,IREALCARE3,IREALCARE4,IREALCARE5} can be used in this problem for follow-up research.

2. The number of signal categories included in the classified signal set studied in this paper is small, and only linear non-memory digital modulation systems are considered. In the future, coded modulation schemes, such as Continuous Phase Modulation (CPM) \cite{codedcpm2,codedcpm3} and trellis coded CPM \cite{codedcpm1,codedcpm4,codedcpm5,codedcpm6}, and modulation combined with different coding schemes used in some practical systems \cite{b3a,b3b,b3c,distributedRaptor,Raptor_ML,JNCC,RCRC,NC1,NC2,NC3,NC4,WRN,cellular1,cellular2,cellular3,MIMO_capacity,network_capacity,UAVdownlink,UAV_THz,UAV_2,RF_energy1,RF_energy2,RF_energy3,RF_energy4}, can also be considered. For a richer set of classified signals, how the modulation recognition performance of various modulation recognition technologies based on deep learning will perform is also a direction worthy of further research.

 3. This paper does not consider the impact of channel environment changes on the modulation classification algorithm of the deep learning model. The impact of channel environment changes on the recognition accuracy can be studied later.




\begin{thebibliography}{99}  
\bibitem{dobre2007survey}O. A. Dobre, A. Abdi, Y. Bar-Ness, and W. Su, “Survey of automatic modulation classification techniques: Classical approaches and new trends,” IET Commun., vol. 1, no. 2, pp. 137–156, Apr. 2007.
\bibitem{wei2000maximum}W. Wei and J. M. Mendel, “Maximum-likelihood classiﬁcation for digital amplitude-phase modulations,” IEEE Trans. Commun., vol. 48, pp. 189–193, 2000.
\bibitem{chavali2011maximum}  V. G. Chavali and C. R. C. M. da Silva, "Maximum-Likelihood Classification of Digital Amplitude-Phase Modulated Signals in Flat Fading Non-Gaussian Channels," in IEEE Transactions on Communications, vol. 59, no. 8, pp. 2051-2056, August 2011, doi: 10.1109/TCOMM.2011.051711.100184.
\bibitem{azzouz1995automatic}  E. E. Azzouz, A. K. Nandi, Automatic identification of digital modulation types. Signal Process.47(1), 55–69 (1995).
\bibitem{liu2012novel}Jian Liu and Qiang Luo, "A novel modulation classification algorithm based on daubechies5 wavelet and fractional fourier transform in cognitive radio," 2012 IEEE 14th International Conference on Communication Technology, Chengdu, 2012, pp. 115-120, doi: 10.1109/ICCT.2012.6511199.
\bibitem{das2016cumulant}D. Das, P. K. Bora and R. Bhattacharjee, "Cumulant based automatic modulation classification of QPSK, OQPSK, 8-PSK and 16-PSK," 2016 8th International Conference on Communication Systems and Networks (COMSNETS), Bangalore, 2016, pp. 1-5, doi: 10.1109/COMSNETS.2016.7439996.
\bibitem{hassan2009automatic}  K. Hassan, I. Dayoub, W. Hamouda and M. Berbineau, "Automatic modulation recognition using wavelet transform and neural network," 2009 9th International Conference on Intelligent Transport Systems Telecommunications, (ITST), Lille, 2009, pp. 234-238, doi: 10.1109/ITST.2009.5399351.
\bibitem{orlic2010multipath}  V. D. Orlic and M. L. Dukic, "Multipath channel estimation algorithm for automatic modulation classification using sixth-order cumulants," in Electronics Letters, vol. 46, no. 19, pp. 1348-1349, 16 Sept. 2010, doi: 10.1049/el.2010.1893.
\bibitem{o2016convolutional}T. J. O’Shea, J. Corgan and T. C. Clancy, "Convolutional radio modulation recognition networks", International conference on engineering applications of neural networks, pp. 213-226, 2016.
\bibitem{zhang2018automatic1}D. Zhang, W. Ding, B. Zhang, C. Xie, H. Li, C. Liu, et al., "Automatic modulation classification based on deep learning for unmanned aerial vehicles", Sensors, vol. 18, no. 3, pp. 924, 2018.
\bibitem{zhang2018automatic2}  M. Zhang, Y. Zeng, Z. Han and Y. Gong, "Automatic modulation recognition using deep learning architectures", 2018 IEEE 19th International Workshop on Signal Processing Advances in Wireless Communications (SPAWC), pp. 1-5, 2018.
\bibitem{mendis2016deep}  G. J. Mendis, J. Wei and A. Madanayake, "Deep learning-based automated modulation classification for cognitive radio", Communication Systems (ICCS) 2016 IEEE International Conference on, pp. 1-6, 2016.
\bibitem{dai2016automatic}A. Dai, H. Zhang and H. Sun, "Automatic modulation classification using stacked sparse auto-encoders", Signal Processing (ICSP) 2016 IEEE 13th International Conference on, pp. 248-252, 2016.

\bibitem{yin2019co}Z. Yin, R. Zhang, Z. Wu and X. Zhang, "Co-Channel Multi-Signal Modulation Classification Based on Convolution Neural Network," 2019 IEEE 89th Vehicular Technology Conference (VTC2019-Spring), Kuala Lumpur, Malaysia, 2019, pp. 1-5, doi: 10.1109/VTCSpring.2019.8746292.
\bibitem{sun2018automatic}  J. Sun, G. Wang, Z. Lin, S. G. Razul and X. Lai, "Automatic Modulation Classification of Cochannel Signals using Deep Learning," 2018 IEEE 23rd International Conference on Digital Signal Processing (DSP), Shanghai, China, 2018, pp. 1-5, doi: 10.1109/ICDSP.2018.8631682.
\bibitem{zhou2019blind}  S. Zhou, Z. Wu, Z. Yin, R. Zhang and Z. Yang, "Blind Modulation Classification for Overlapped Co-Channel Signals Using Capsule Networks," in IEEE Communications Letters, vol. 23, no. 10, pp. 1849-1852, Oct. 2019, doi: 10.1109/LCOMM.2019.2929799.
\bibitem{lin2017deep}X. Lin, R. Liu, et al, "A Deep Convolutional Network Demodulator for Mixed Signals with Different Modulation Types," 2017 IEEE 15th Intl Conf on Dependable, Autonomic and Secure Computing, Orlando, FL, 2017, pp. 893-896, doi: 10.1109/DASC-PICom-DataCom-CyberSciTec.2017.150.




\bibitem{o2016radio}T. J. O'Shea, N. West. Radio machine learning dataset generation with gnu radio[C].Proceedings of the GNU Radio Conference, 2016.
\bibitem{o2017semi}  T. J. O'Shea, N. West, M. Vondal, et al. Semi-supervised radio signal identification[C]. IEEE International Conference on Advanced Communication Technology, 2017, 33-38.
\bibitem{west2017deep}  N. E. West, T. J. O'Shea. Deep architectures for modulation recognition[C]. IEEE International Symposium on Dynamic Spectrum Access Networks, 2017, 1-6.


\bibitem{hu2018robust}S. Hu, Y. Pei, P. P. Liang, et al. Robust Modulation Classification under Uncertain Noise Condition Using Recurrent Neural Network[C]. IEEE Global Communications Conference, 2018, 1-7.
\bibitem{meng2018automatic}F. Meng, P. Chen, L. Wu, et al. Automatic modulation classification: A deep learning enabled approach[J]. IEEE Transactions on Vehicular Technology, 2018, 67(11): 10760-10772.


\bibitem{zhu2015automatic}Z. Zhu, A. K. Nandi. Automatic modulation classification: principles, algorithms and applications[M]. John Wiley \& Sons, 2014.
\bibitem{swami2000hierarchical}A. Swami, B. M Sadler. Hierarchical digital modulation classification using cumulants[J].IEEE Transactions on Communications, 2000, 48(3): 416-429.
\bibitem{peng2017modulation} S. Peng, H. Jiang, H. Wang, et al. Modulation classification using convolutional Neural Network based deep learning model[C]. IEEE Wireless and Optical Communication Conference, 2017, 1-5. 


\bibitem{IREALCARE1}L. Meng, K. Ge, Y. Song, D. Yang, and Z. Lin, "Long-term wearable electrocardiogram signal monitoring and analysis based on convolutional neural network", the IEEE Transactions on Instrumentation \& Measurement, Vol. 70, April 2021, DOI:10.1109/TIM.2021.3072144

\bibitem{IREALCARE2} Z. Chen, Z. Lin, P. Wang, and M. Ding, Negative-ResNet: noisy ambulatory electrocardiogram signal classification scheme, Neural Computing and Applications, Vol. 33, Issue 14, July 2021, pp.  8857-8869

\bibitem{IREALCARE3} P, Wang, Z. Lin, Z. Chen, X. Yan, M. Ding, A Wearable ECG Monitor for Deep Learning-Based Real-Time Cardiovascular Disease Detection, https://arxiv.org/abs/2201.10083

\bibitem{IREALCARE4} X. Yan, Z. Lin, P. Wang, Wireless Electrocardiograph Monitoring Based on Wavelet Convolutional Neural Network, Proceedings of the IEEE WCNC 2020.

\bibitem{IREALCARE5} M. Liu, Z. Lin, P. Xiao, and W. Xiang. "Human Biometric Signals Monitoring based on WiFi Channel State Information using Deep Learning." arXiv preprint arXiv:2203.03980 (2022).


\bibitem{codedcpm2}Z. Lin and T. Aulin, “On Combined Ring Convolutional Coded Quantization and CPM for Joint Source and Channel Coding”, Transactions on Emerging Telecommunications Technologies, Special Issue on ’New Directions in Information Theory’, Vol.19, No.4. June 2008, pp. 443-453. 

\bibitem{codedcpm3} Z. Lin and T. Aulin, “Joint Source-Channel Coding using Combined TCQ/CPM: Iterative Decoding”, IEEE Transactions on Communications, VOL.53, NO. 12, Dec. 2005, pp. 1991-1995.


\bibitem{codedcpm1}Z. Lin and T. Aulin, “Joint Source and Channel Coding using Punctured Ring Convolutional Coded CPM”, IEEE Transactions on Communications, Vol. 56, No. 5, May, 2007, pp. 712-723. 

\bibitem{codedcpm4}	Z. Lin and B. Vucetic, “Performance Analysis on Ring Convolutional Coded CPM”, IEEE Transactions on Wireless Communications, Vol. 8, No. 9, Sept. 2009, pp. 4848-4854. The latest impact factor is 4.951
\bibitem{codedcpm5}	Z. Lin and B. Vucetic, “Spatial Frequency Scheduling for SC-FDMA Based Uplink Multi-user MIMO Systems”, IET Communications, V. 3, No. 7, July, 2009, pp. 163-165.

\bibitem{codedcpm6}	Z. Lin and T. Aulin, “On Joint Source and Channel Coding using trellis coded CPM: Analytical Bounds on the Channel Distortion”, IEEE Transactions on Information Theory, Vol. 53, No. 13, Sept. 2007. pp. 3081-3094. The latest impact factor is 2.679



\bibitem{b3a}	Y. Hu, P. Wang, Z. Lin, M. Ding, YC. Liang, “Performance Analysis of Ambient Backscatter Systems with LDPC-coded Source Signals” in IEEE Transactions on Vehicular Technology, vol. 70, no. 8, pp. 7870-7884, Aug. 2021, doi: 10.1109/TVT.2021.3093912.

\bibitem{b3b}	Y. Hu, P. Wang, Z. Lin, M. Ding, YC. Liang, “Machine Learning Based Signal Detection for Ambient Backscatter Communications”, 2019 IEEE International Conference on Communications (ICC): SAC Internet of Things Track. 

\bibitem{b3c} Z. Xing, Z. Lin, M. Ding,  “Outage Capacity Analysis for Ambient Backscatter Communication Systems”, 2018 28th International Telecommunication Networks and Applications Conference (ITNAC).

\bibitem{distributedRaptor}J. Yue, Z. Lin, B. Vucetic, G. Mao, T. Aulin, "Performance analysis of distributed raptor codes in wireless sensor networks", IEEE Transactions on Communications 61 (10), 2013, 4357-4368

\bibitem{Raptor_ML}P. Wang, G. Mao, Z. Lin, M Ding, W. Liang, X. Ge, Z. Lin, "Performance analysis of raptor codes under maximum likelihood decoding", IEEE Transactions on Communications 64 (3), 2016, 906-917

\bibitem{JNCC}K. Pang, Z. Lin, Y. Li, B. Vucetic, "Joint network-channel code design for real wireless relay networks", the 6th International Symposium on Turbo Codes \& Iterative Information, 2010, 429-433.

\bibitem{RCRC}Z. Lin, A. Svensson, "New rate-compatible repetition convolutional codes",
IEEE Transactions on Information Theory 46 (7),  2651-2659


\bibitem{NC1}Z Lin, B Vucetic, "Power and rate adaptation for wireless network coding with opportunistic scheduling", 2008 IEEE International Symposium on Information Theory, 21-25

\bibitem{NC2} T Ding, M Ding, G Mao, Z Lin, AY Zomaya, D López-Pérez, "Performance analysis of dense small cell networks with dynamic TDD", IEEE Transactions on Vehicular Technology 67 (10), 9816-9830, 2018.

\bibitem{NC3}P Wang, G Mao, Z Lin, X Ge, BDO Anderson, "Network coding based wireless broadcast with performance guarantee", IEEE Transactions on Wireless Communications 14 (1), 532-544, 2014.

\bibitem{NC4}K Pang, Z Lin, Y Li, B Vucetic,"Distributed network-channel codes design with short cycles removal", IEEE Wireless Communications Letters 2 (1), 62-65, 2012.

\bibitem{WRN}J. Yue; Z. Lin; B. Vucetic; G. Mao; M. Xiao; B. Bai; K. Pang, "Network Code Division Multiplexing for Wireless Relay Networks,"  IEEE Transactions on Wireless Communications, vol.14, no.10, pp.5736-5749, Oct. 2015.

\bibitem{cellular1}Z. Lin, P. Xiao and B. Vucetic, “Analysis of Receiver Algorithms for LTE SC-FDMA Based Uplink MIMO Systems”, IEEE Transactions on Wireless Communications, Vol. 9, No. 1, Nov. 2010, pp. 60-65. 

\bibitem{cellular2}Y. Chen, M. Ding, D. Lopez-Perez, J. Li, Z. Lin, B. Vucetic, "Dynamic reuse of unlicensed spectrum: An inter-working of LTE and WiFi", IEEE Wireless Communications 24 (5), 52-59

\bibitem{cellular3}Y Chen, J Li, Z Lin, G Mao, B Vucetic, "User association with unequal user priorities in heterogeneous cellular networks", IEEE Transactions on Vehicular Technology 65 (9), 7374-7388

\bibitem{MIMO_capacity}Z. Lin, B. Vucetic, J. Mao, "Ergodic capacity of LTE downlink multiuser MIMO systems", 2008 IEEE International Conference on Communications, 3345-3349.

\bibitem{network_capacity}G. Mao, Z. Lin, X. Ge, Y. Yang, "Towards a simple relationship to estimate the capacity of static and mobile wireless networks", IEEE transactions on wireless communications 12 (8), 2014, 3883-3895	

\bibitem{UAVdownlink}D López-Pérez, M Ding, H Li, LG Giordano, G Geraci, A Garcia-Rodriguez, Z. Lin, M. Hassan, "On the downlink performance of UAV communications in dense cellular networks", 2018 IEEE global communications conference (GLOBECOM), 1-7

\bibitem{UAV_THz}	X. Wang, P. Wang, M. Ding, Z. Lin, L. Hanzo and B. Vucetic, “Performance Analysis of TeraHertz Unmanned Aerial Vehicular Networks”, in IEEE Transactions on Vehicular Technology, vol. 69, no. 12, pp. 16330-16335, Dec. 2020, doi: 10.1109/TVT.2020.3035831. 

\bibitem{UAV_2}C Liu, M Ding, C Ma, Q Li, Z Lin, YC Liang, "Performance analysis for practical unmanned aerial vehicle networks with LoS/NLoS transmissions", IEEE International Conference on Communications Workshops (ICC Workshops), 2018, 1-6.

\bibitem{RF_energy1}J. Wang, B. Li, G. Wang, Z. Lin, H. Wang, and G. Chen, Optimal Power Splitting for MIMO SWIPT Relaying Systems with Direct Link in IoT Networks, Physical Communication, Volume 43, December 2020. 

\bibitem{RF_energy2}D. Zhai, H. Chen, Z. Lin. Y. Li and B. Vucetic, “Accumulate Then Transmit: Multi-user Scheduling in Full-Duplex Wireless-Powered IoT Systems”, IEEE Internet of Things Journal, Volume: 5 , Issue: 4 , Aug. 2018. 

\bibitem{RF_energy3} H. Chen, Y. Ma. Z. Lin, Y. Li and B. Vucetic, “Distributed Power Control in Interference Channels with QoS Constraints and RF Energy Harvesting: A Game-Theoretic Approach”, IEEE Transactions on Vehicular Technology, Volume: 65, Issue: 12, Dec. 2016. pp.10063 – 10069. 

\bibitem{RF_energy4} Y. Ma, H. Chen, Z. Lin, Y. Li and B. Vucetic, "Distributed and Optimal Resource Allocation for Power Beacon-Assisted Wireless-Powered Communications,", IEEE Transactions on  Communications, vol.63, no.10, pp.3569-3583, Oct. 2015.















\end{thebibliography}
\end{document}